\documentclass[structabstract]{aa}  
\usepackage{graphicx}
\usepackage{color}
\usepackage{epsfig}
\usepackage{supertabular}
\usepackage{longtable}
\usepackage{subfigure}
\usepackage{txfonts}
\usepackage{natbib}
\bibpunct{(}{)}{;}{a}{}{,}
\newcommand{\vrad}{v_{\mathrm{rad}}}
\newcommand{\kms}{km~s$^{-1}$}
\usepackage{booktabs}
\begin{document}
   \title{A large population of ultra-compact dwarf galaxies \\ in the Hydra\,I cluster\thanks{Based on observations obtained at the European Southern Observatory, Chile (Observing Programmes 082.B-0680 and 076.B-0154).}}

   %\subtitle{}

   \author{I. Misgeld\inst{1,3} \and S. Mieske\inst{2} \and M. Hilker\inst{3} \and T. Richtler \inst{4}  \and I. Y. Georgiev \inst{5} \and Y. Schuberth \inst{5}}

   %\offprints{}

   \institute{
         Universit\"ats-Sternwarte M\"unchen, Scheinerstr. 1, 81679 M\"unchen, Germany \\
         \email{imisgeld@mpe.mpg.de}   
    	\and European Southern Observatory, Alonso de Cordova 3107, Vitacura, Santiago, Chile \\
	         \email{smieske@eso.org}
     	\and European Southern Observatory, Karl-Schwarzschild-Strasse 2, 85748 Garching bei M\"unchen, Germany \\
         \email{mhilker@eso.org}	
       	\and Universidad de Concepci\'on, Departamento de Astronom\'ia, Casilla 160-C, Concepci\'on, Chile
	    \and Argelander-Institut f\"ur Astronomie, Universit\"at Bonn, Auf dem H\"ugel 71, 53121 Bonn}

   \date{}

% \abstract{}{}{}{}{} 
% 5 {} token are mandatory
 
  \abstract
   {} %leave it empty if necessary  
   {We performed a large spectroscopic survey of compact, unresolved objects in the core of the Hydra\,I galaxy cluster (Abell 1060), with the aim of identifying ultra-compact dwarf galaxies (UCDs), and investigating the properties of the globular cluster (GC) system around the central cD galaxy NGC~3311.}
   {We obtained VIMOS medium resolution spectra of about 1200 candidate objects with apparent magnitudes $18.5 < V < 24.0$~mag, covering both the bright end of the GC luminosity function and the luminosity range of all known UCDs.}
   {By means of spectroscopic redshift measurements, we identified 118 cluster members, from which 52 are brighter than $M_V=-11.0$~mag, and can therefore be termed UCDs. The brightest UCD in our sample has an absolute magnitude of $M_V=-13.4$~mag (corresponding to a mass of $\gtrsim 5\times 10^7$~M$_{\sun}$) and a half-light radius of 25~pc. This places it among the brightest and most massive UCDs ever discovered. Most of the GCs/UCDs are both spatially and dynamically associated to the central cD galaxy. The overall velocity dispersion of the GCs/UCDs is comparable to what is found for the cluster galaxies. However, when splitting the sample into a bright and a faint part, we observe a lower velocity dispersion for the bright UCDs/GCs than for the fainter objects. At a dividing magnitude of $M_V=-10.75$~mag, the dispersions differ by more than 200~\kms, and up to 300~\kms~for objects within $5\arcmin$ around NGC~3311.}
   {We interpret these results in the context of different UCD formation channels, and conclude that interaction driven formation seems to play an important role in the centre of Hydra\,I.}


   \maketitle 
%
%________________________________________________________________

\section{Introduction}
Within the last decade, a new class of stellar systems, called "ultra-compact dwarf galaxies" (UCDs), has been discovered in the nearby galaxy clusters Fornax, Virgo, Centaurus and Coma \citep[e.g.][]{1999A&AS..134...75H, 2000PASA...17..227D, 2003Natur.423..519D, 2001ApJ...560..201P, 2004A&A...418..445M, 2007A&A...472..111M, 2009A&A...498..705M, 2005ApJ...627..203H, 2006AJ....131..312J, 2007MNRAS.382.1342F, 2009AJ....137..498G, 2009MNRAS.397.1816P, 2010arXiv1009.3950C, 2010ApJ...722.1707M}. Recently, UCDs have also been identified in several group environments \citep[e.g.][]{2007MNRAS.378.1036E, 2007A&A...469..147R, 2008AJ....136.2295B, 2009MNRAS.394L..97H, 2011A&A...525A..86D, 2011arXiv1102.0001N}. They are characterised by evolved stellar populations \citep[e.g.][]{2006AJ....131.2442M, 2007AJ....133.1722E}, typical luminosities of $-13.5<M_V<-11.0$~mag, masses of $2\times 10^6 < m < 10^8$~M$_{\sun}$ \citep[e.g.][]{2008A&A...487..921M}, and half-light radii of $10<r_{\mathrm{h}}<100$~pc. Unlike globular clusters (GCs), UCDs follow a luminosity-size relation, in the sense that more luminous UCDs have larger half-light radii \citep[e.g.][]{2005ApJ...627..203H, 2008AJ....136..461E}. Moreover, UCDs show enhanced dynamical mass-to-light ($M/L$) ratios in comparison to Galactic globular clusters of similar metallicity \citep[e.g.][]{2005ApJ...627..203H, 2007A&A...463..119H, 2007A&A...469..147R, 2008MNRAS.386..864D, 2009MNRAS.394.1529D, 2008A&A...487..921M, 2010ApJ...712.1191T}.

UCDs are hence of intermediate nature between dwarf elliptical galaxies and GCs, however, they are not a homogeneous class of objects. For example, Virgo UCDs are on average more metal-poor, have larger $\alpha$ abundances, and extend to higher $M/L$ ratios than UCDs in Fornax \citep{2005ApJ...627..203H, 2006AJ....131.2442M, 2007A&A...472..111M, 2007AJ....133.1722E, 2007A&A...463..119H}, although this might not be true if one could examine complete samples in either cluster. The colours of UCDs cover the full range of colours observed for regular GCs \citep[e.g.][]{2008AJ....136..461E, 2009gcgg.book...51H}, but the brightest UCDs tend to have red colours. Apparently, they represent the extension of the red (metal-rich) GC population to higher luminosities. Blue (metal-poor) UCDs, on the other hand, share the location of nuclei of early-type dwarf galaxies in a colour--magnitude diagram and follow a mass-colour relation \citep[the "blue tilt", e.g.][and references therein]{2010ApJ...710.1672M, 2011arXiv1102.0001N}. These properties may indicate that more than one formation channel for UCDs exists.

As one possible formation scenario, it was proposed that UCDs formed from the amalgamation of many young, massive star clusters during the interaction of gas-rich galaxies \citep[e.g.][]{1998MNRAS.300..200K, 2002MNRAS.330..642F}. Indeed, hundreds of young, massive star clusters have been discovered in the merging Antennae galaxies NGC~4038/4039 \citep[e.g.][]{1999AJ....118.1551W}. They are themselves clustered into groups, and will likely merge on time scales of a few tens to a hundred Myr. An example of such a merged star cluster complex could be the massive cluster W3 in NGC~7252, which has, apart from its young age of 500~Myr, properties very similar to those of UCDs \citep{2004A&A...416..467M, 2005MNRAS.359..223F}. Thus, UCDs might be regarded as the brightest and most massive (metal-rich) globular clusters, representing the bright tail of the globular cluster luminosity function (GCLF) \citep[e.g.][]{2002A&A...383..823M, 2004A&A...418..445M}. This scenario is supported by the smooth appearance of the GCLF of several giant elliptical galaxies, which extends continuously to very bright objects which fall into the luminosity range of UCDs \citep[e.g.][]{2005A&A...438..103M, 2008ApJ...681.1233W, 2009ApJ...699..254H}. In this context, the high $M/L$ ratios of the most massive UCDs can be interpreted as either the consequence of a non-canonical IMF, top-heavy \citep{2009MNRAS.394.1529D} or bottom-heavy \citep{2008ApJ...677..276M}, or as the consequence of GC/UCD formation in dark matter haloes \citep[e.g.][]{2008MNRAS.391..942B}. In a recent study by \citet{2010MNRAS.408L..16G}, it was shown that also the size-luminosity relation of UCDs can be explained by assuming that low-mass GCs have formed with the same relation as the more massive UCDs, and have moved away from this relation due to dynamical evolution.

An alternative suggestion is that UCDs are genuine compact dwarf galaxies, originating from primordial small scale dark matter peaks and having survived the galaxy cluster formation and evolution until the present time \citep{2001ApJ...560..201P, 2004PASA...21..375D}.

Another possible formation scenario is that UCDs are the remnant nuclei of dwarf galaxies which lost their outer envelopes during the interaction with the tidal field of the parent galaxy or galaxy cluster \citep[e.g.][]{1994ApJ...431..634B, 2001ApJ...552L.105B, 2003MNRAS.344..399B, 2008MNRAS.385.2136G}. Observations show that the structural parameters of the brightest, metal-poor UCDs resemble present-day nuclei of dwarf galaxies. However, the high metallicities of many Fornax UCDs seem to be contradictory to this formation scenario \citep{2006AJ....131.2442M, 2007A&A...472..111M}.

\begin{table}
	\caption{Observing Log. The first column indicates the pointing and the mask number, columns 2 and 3 give the central coordinates of the pointing (cf. Fig.~\ref{fig:fields}), column 4 lists the date of observation, and column 5 the exposure time.}
	\label{tab:obslog}
	\centering	
		\begin{tabular}{l c c c c}
		\hline\hline
		Pointing/ & $\alpha$(2000.0) & $\delta$(2000.0) & Date & $T_{exp}$ \\
		Mask &  [h:m:s] & [$^\circ$:$\arcmin$:$\arcsec$] &  & [sec]\\
		\hline
		P1M1 & 10:36:15.7 & -27:37:12.0 & 2007/11/11 & $2\times 2100$ \\
		P1M2 & 10:36:15.7 & -27:37:12.0 & 2008/12/26 & $2\times 2555$ \\		
		P2M1 & 10:36:58.6 & -27:27:40.4 & 2008/12/27 & $2\times 2555$ \\		
		P2M2 & 10:36:58.6 & -27:27:40.4 & 2009/01/31 & $2\times 2555$ \\		
		P3M1 & 10:36:07.1 & -27:24:29.9 & 2009/02/18 & $2\times 2555$ \\		
		P3M2 & 10:36:07.1 & -27:24:29.9 & 2009/02/01 & $2\times 2555$ \\		
		P4M1 & 10:37:09.8 & -27:41:11.3 & 2009/02/28 & $2\times 2555$ \\		
		P4M2 & 10:37:09.8 & -27:41:11.3 & 2009/02/24 & $2\times 2555$ \\		
		P5M1 & 10:36:18.0 & -27:28:39.6 & 2006/02/28 & $4\times 2185$ \\		
		P5M2 & 10:36:18.0 & -27:28:39.6 & 2007/01/27 & $4\times 2185$ \\		
		P5M3 & 10:36:18.0 & -27:28:39.6 & 2007/02/22 & $4\times 2185$ \\		
		P6M1 & 10:36:28.9 & -27:30:58.7 & 2007/01/21 & $4\times 2185$ \\		
		P6M2 & 10:36:28.9 & -27:30:58.7 & 2007/01/24 & $4\times 2185$ \\		
		\hline
		\end{tabular}
\end{table}

Because of the apparent heterogeneity of UCDs, it is essential to broaden the environmental baseline of UCD studies beyond the well studied clusters Fornax and Virgo. In this paper, we present a spectroscopic census of compact objects in the core region of the Hydra\,I galaxy cluster (Abell 1060). Hydra\,I is well suited for a search for UCDs, since the cluster centre is dominated by the prominent cD galaxy NGC~3311, which exhibits a very pronounced diffuse light component and an extremely rich globular cluster system \citep{1977ApJ...212..317V, 1995AJ....109.1033M, 2005A&A...438..103M, 2008ApJ...681.1233W}. Our aim is to investigate the bright end of the GCLF, where UCDs are expected to be found, and the globular cluster system of the two central cluster galaxies NGC~3311 and NGC~3309. For this, we analyse two spectroscopic surveys, which were carried out with the VIsible MultiObject Spectrograph \citep[VIMOS,][]{2003SPIE.4841.1670L} mounted on UT3 at the VLT. One survey explicitly targets at UCD candidates (ESO observing programme 082.B-0680, PI: I.~Misgeld), the other one targets at fainter sources, mainly GC candidates (ESO observing programme 076.B-0154, PI: T.~Richtler). The significance of the data presented here for the dynamics of NGC~3311 is discussed in a parallel contribution \citep{2011arXiv1103.2053R}.

\begin{figure}
	\resizebox{\hsize}{!}{\includegraphics{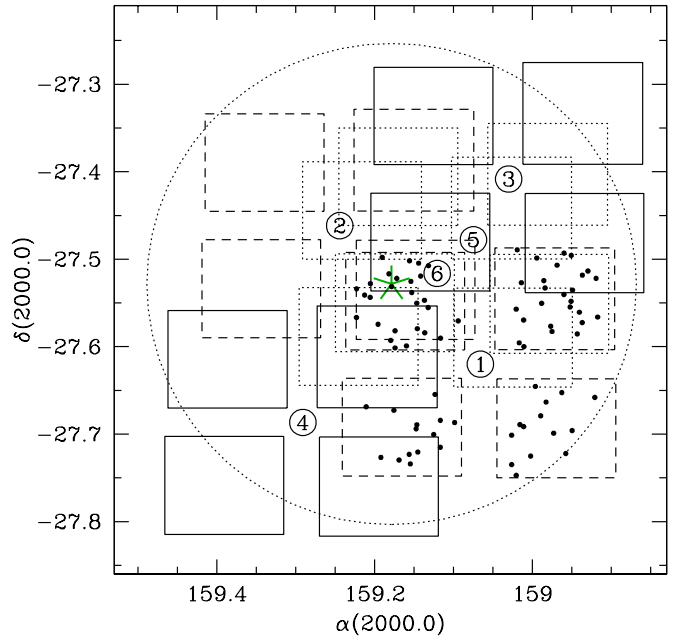}}
	\caption{Map of the surveyed area in the Hydra\,I cluster. The dashed squares mark the two pointings, which had been approved for observations in 2007 (Nr.~1 and Nr.~2), but from which only pointing Nr.~1 was observed with one slit mask. Black dots are all compact sources observed in this particular programme \citep[see][]{2008A&A...486..697M}. The solid squares mark the two additional pointings for the new UCD survey in 2009 (Nr.~3 and Nr.~4). The two pointings of the GC survey (see Sect.~\ref{sec:GCsample}) are represented by the dotted squares (Nr.~5 and Nr.~6). The large dotted circle indicates the projected cluster core-radius of $r_{\mathrm{c}}\sim 170\ h^{-1}$~kpc \citep{1995ApJ...438..527G}, adopting $h=0.75$. The large green asterisk marks the location of the central cluster galaxy NGC~3311.}
	\label{fig:fields}
\end{figure}

Throughout this paper we adopt a Hydra\,I distance modulus of $(m-M)=33.37$~mag, which is the mean value from different studies \citep[see][and references therein]{2005A&A...438..103M}. This corresponds to a physical scale of 229~pc/arcsec at $47.2$~Mpc.

\begin{figure*}
	\centering
	\subfigure{\includegraphics[width=6cm]{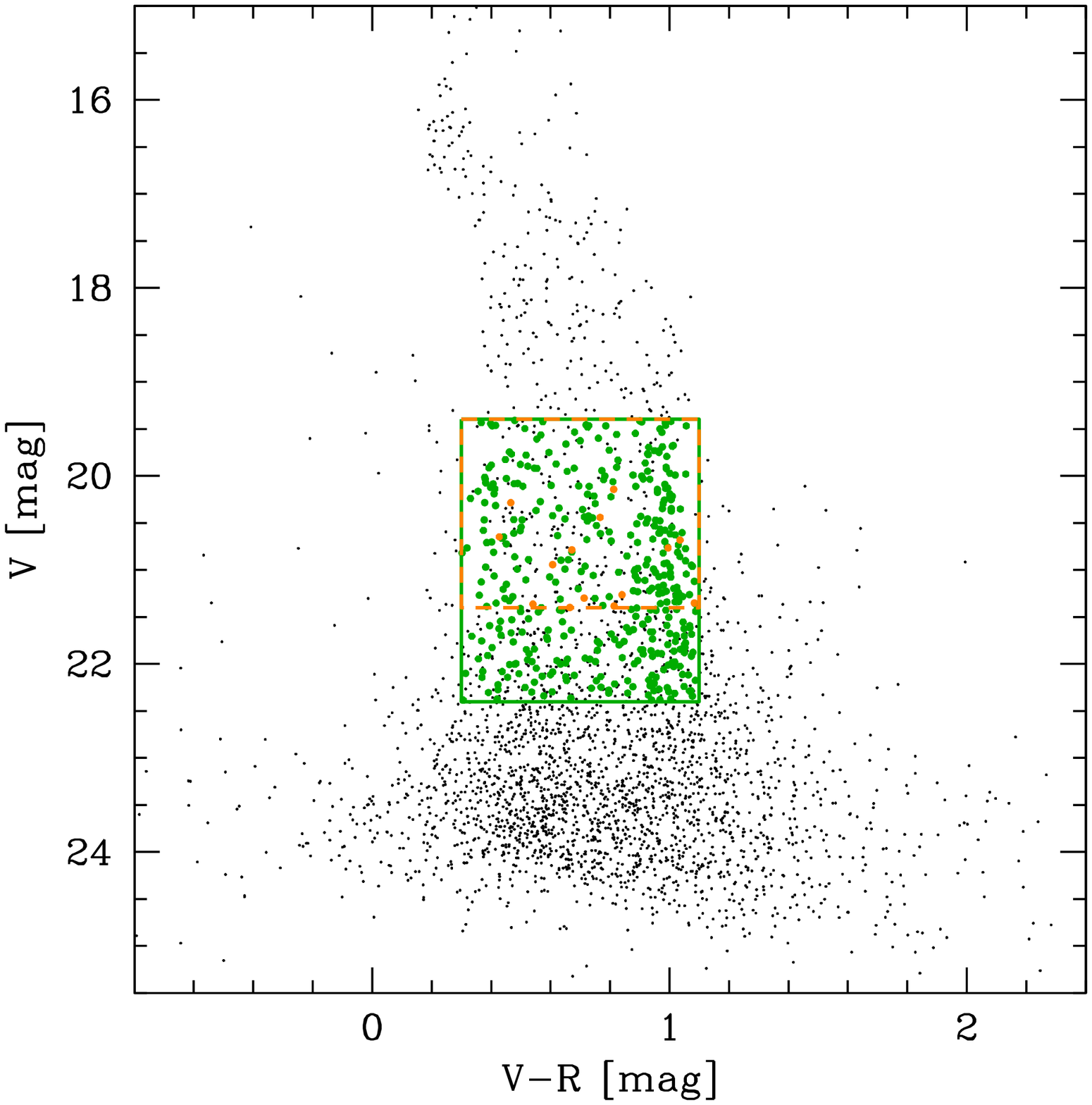}}
	\subfigure{\includegraphics[width=6cm]{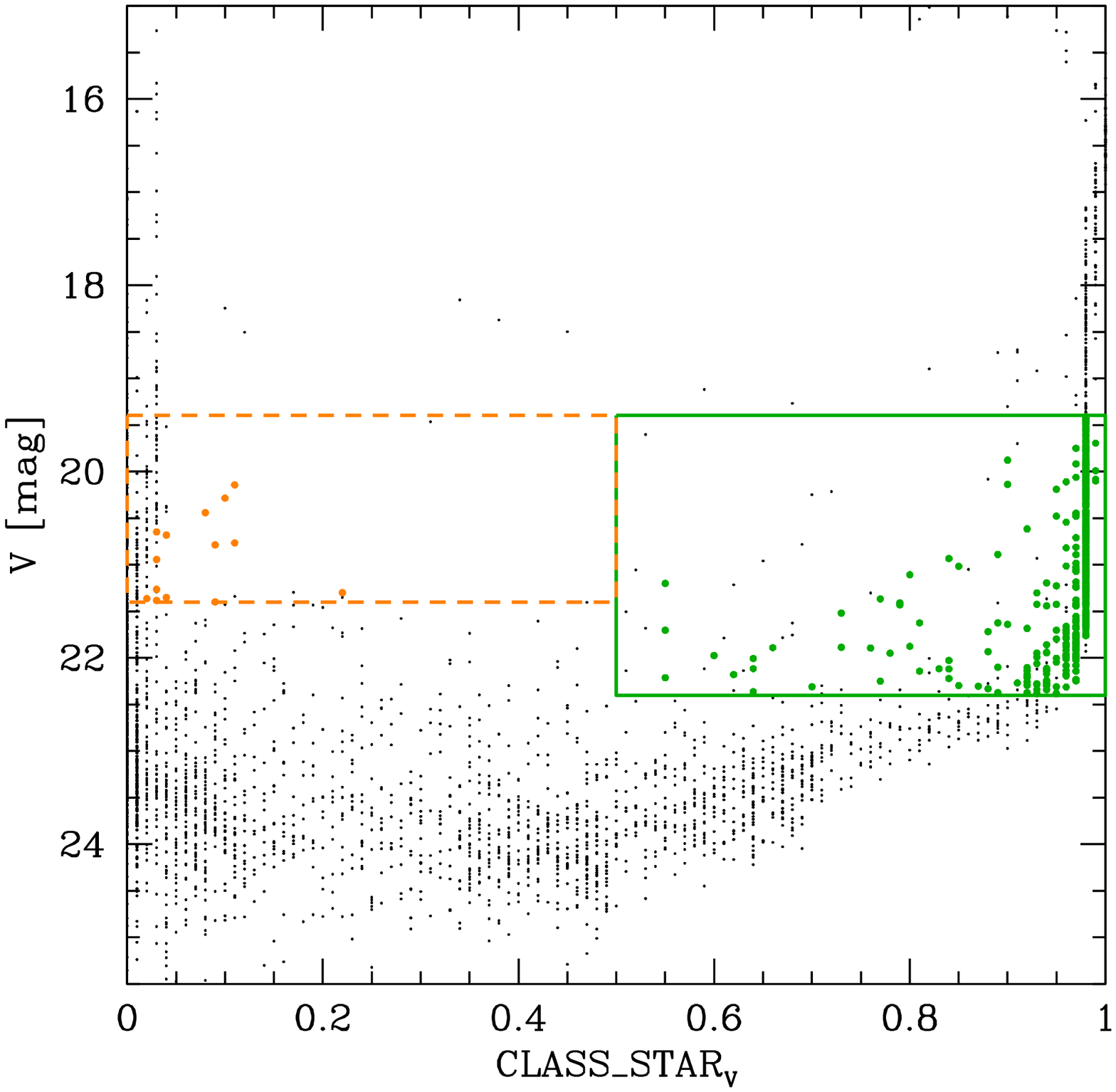}}
	\subfigure{\includegraphics[width=6cm]{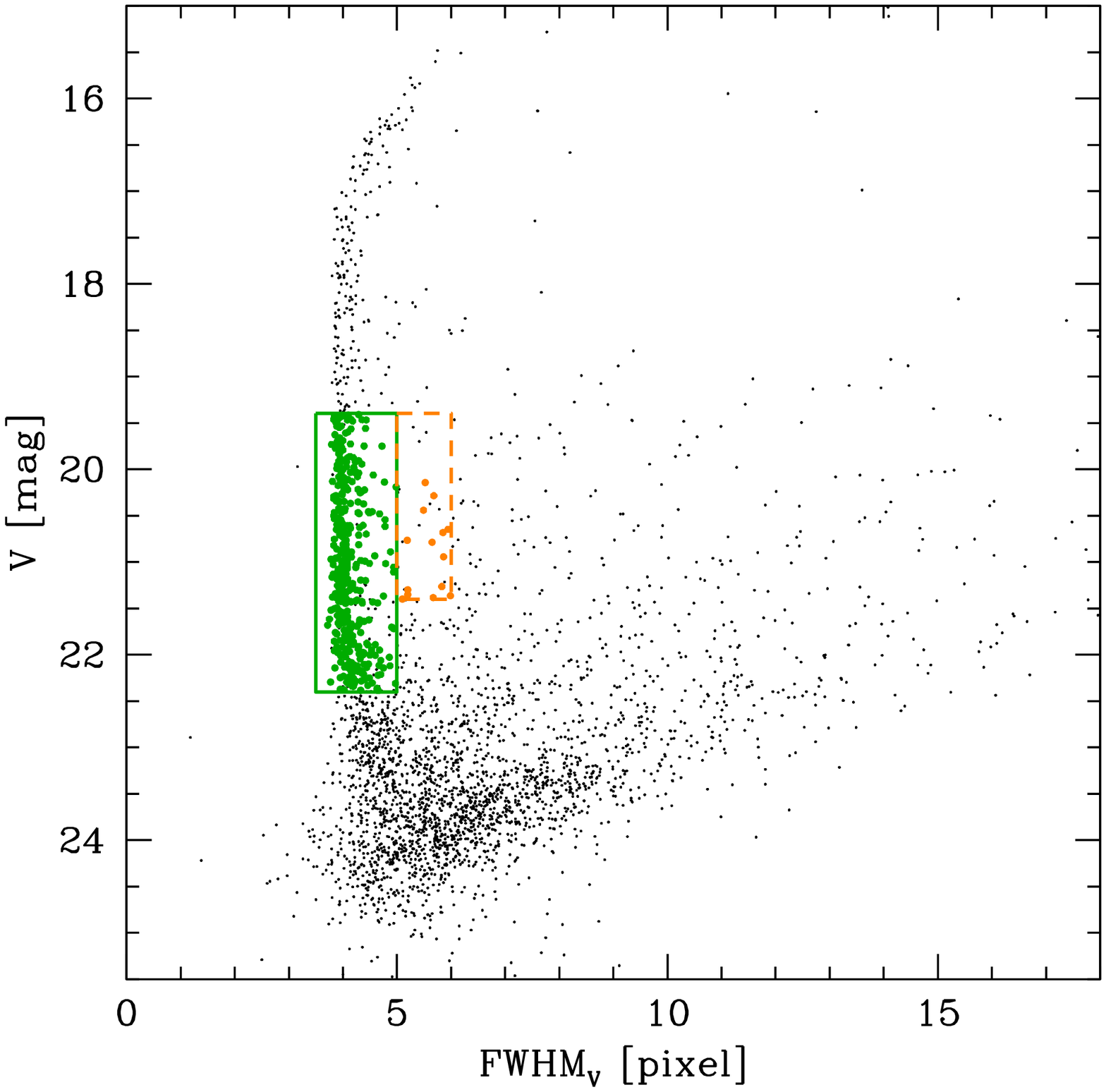}}
	\caption{Selection of the UCD candidates. The apparent $V$ magnitude of all detected objects in pointing Nr.~4 (black dots) is plotted vs. their $V-R$ colour (\textit{left panel}), the SExtractor star-classifier value (\textit{middle panel}), and their FWHM (\textit{right panel}). Green solid boxes represent the selection criteria for the main UCD sample, orange dashed boxes are the cuts set for the additional sub-sample of slightly resolved UCD candidates (see text for details). Green and orange points are those objects which fulfil the selection criteria.}
	\label{fig:ucdselection}
\end{figure*}

\section{Observations and candidate selection}
\label{sec:candidates}
VIMOS allows the simultaneous observation of 4 quadrants in one telescope pointing. Each quadrant is of dimension $7'\times 8'$, with a gap of about $2'$ between the quadrants. For the UCD survey, we placed four multi-object spectroscopy (MOS) pointings around NGC~3311, the central cD galaxy of the Hydra\,I cluster (pointings Nr.~1--4 in Fig.~\ref{fig:fields}). Each MOS pointing was observed with two different slit masks which were created with the VIMOS mask creation software VMMPS. Pointing Nr.~1 was already observed with one slit mask in 2007 as part of our previous observing programme 076.B-0293 (see \cite{2008A&A...486..697M} for details). For reasons of consistency, we include the results from this run into the following analyses. For the new UCD survey, this pointing was re-observed with one additional slit mask. Table~\ref{tab:obslog} lists all observations analysed in this paper.

\subsection{The UCD survey}
\label{sec:ucdcands}
The candidates for pointings Nr.~1 and Nr.~2 were selected as described in \cite{2008A&A...486..697M}, i.e. being unresolved in the VIMOS $V$- and $R$-band pre-images, and restricted in apparent magnitude and colour to $19.2<V<22.7$~mag and $0.48<V-R<0.93$~mag.

The UCD candidates for the new MOS pointings Nr.~3 and Nr.~4 were again selected from VIMOS pre-imaging in $V$ and $R$, however in a slightly different manner. Figure~\ref{fig:ucdselection} shows exemplarily for pointing Nr.~4 all SExtractor \citep{1996A&AS..117..393B} detections in the four quadrants, and the UCD candidates that were selected by the following requirements:

\begin{enumerate}
 \item Being unresolved, as judged by the SExtractor star-galaxy separator (CLASS\_STAR~$> 0.5$, with 1 for an unresolved source (``star'') and 0 for a resolved source (``galaxy'')). In order to minimize false detections, we additionally set limits on the FWHM, so that the selected candidates were clearly located on the stellar sequence in the magnitude vs. FWHM plot (see right panel of Fig.~\ref{fig:ucdselection}).
 \item Having apparent magnitudes of $19.4<V<22.4$~mag, corresponding to absolute magnitudes of $-14.2<M_V<-11.2$~mag, adopting an interstellar absorption coefficient of $A_V=0.26$~mag \citep{1998ApJ...500..525S}. This encompasses the luminosity range of all known UCDs, although the distinction between UCDs and GCs at the faint magnitude limit is arbitrary to some extent.
 \item Having colours of $0.3<V-R<1.1$~mag. This extends the colour range of previous studies \citep[e.g.][]{2007A&A...472..111M, 2009A&A...498..705M} to both the blue and the red side, and thus prevents a possible bias caused by a too narrow colour window.
\end{enumerate}

\noindent
As shown in \citet{2009A&A...498..705M} for UCD candidates in the Centaurus cluster, the SExtractor star-classifier value flips from $\sim 1$ to $\sim 0$ at a certain physical size (i.e. $r_{\mathrm{eff}}$) of the object. At the Centaurus cluster distance of $\sim 45$~Mpc \citep{2005A&A...438..103M} and a seeing of $\sim 0.8''$ this happens at $r_{\mathrm{eff}}\gtrsim 70$~pc \citep[cf. Fig.~1 in][]{2009A&A...498..705M}. Since the Hydra\,I cluster lies at about the same distance \citep{2005A&A...438..103M}, and the pre-imaging was done under similar seeing conditions, the limitation to objects with CLASS\_STAR~$> 0.5$ would not select the largest known UCDs with effective radii $r_{\mathrm{eff}}\simeq 100$~pc \citep[e.g.][]{2008AJ....136..461E}.

We therefore defined an additional sub-sample of slightly resolved UCD candidates, having brighter magnitudes ($19.4<V<21.4$~mag), lower star-classifier values (CLASS\_STAR~$\leq 0.5$), and a slightly larger FWHM than the primary UCD candidates (see Fig.~\ref{fig:ucdselection}).

\citet{2007ApJ...668L..35W} and \citet{2008ApJ...681.1233W}, W7W8 hereafter, performed a photometric study of the globular cluster system around NGC~3311 and NGC~3309 and identified several UCD candidates. We included 48 objects with magnitudes $i^\prime < 22.2$~mag from their list of candidates to our spectroscopic survey (see also Fig.~\ref{fig:wehnercmd}).

All spectra were obtained with VIMOS, using the medium resolution MR grism and the order sorting filter GG475. The grism gives a wavelength coverage of [4800:10\,000]~\AA~at a dispersion of 2.5~\AA/pixel. The pixel scale is $0.205''$/pixel, so that with a slit width of $1''$ the instrumental resolution corresponds to a $\mathrm{FWHM}\approx5$~pixel, or 12~\AA. This equates to a velocity resolution of 600~\kms~at 6000~\AA.  The average seeing for this set of spectroscopic observations was $\sim 0.8''$. The total on-source integration time was 1.4 hours, subdivided into two exposures.

\subsection{The GC survey}
\label{sec:GCsample}
Two MOS pointings were observed for the GC survey. Pointing Nr.~5 was observed with three different slit masks, pointing Nr.~6 with two (cf. Table~\ref{tab:obslog} and Fig.~\ref{fig:fields}). From the pre-images in $V$- and $R$-band, all unresolved sources (CLASS\_STAR~$> 0.7$) with colours $0.45 < V-R < 0.80$~mag and magnitudes $18.5 < V < 24.0$~mag were selected as GC candidates.

The spectra of the GC candidates were obtained with the same instrument setup as for the UCD candidates, but given the fainter magnitudes of most of the targets, the total on-source integration time was 2.4 hours. The seeing ranged between $0.6''$ and $1.2''$.

\begin{figure}
	\resizebox{\hsize}{!}{\includegraphics{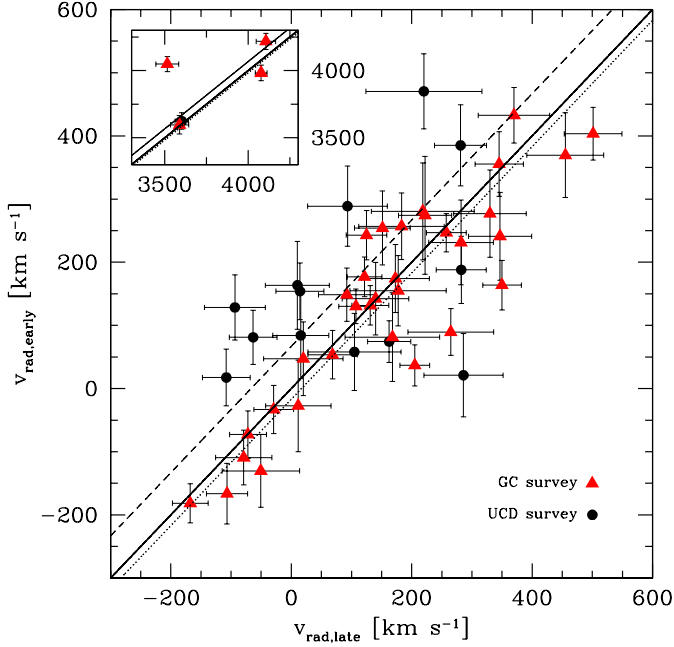}}
	\caption{Comparison of radial velocities for objects which were observed twice, in either the GC survey (red triangles) or the UCD survey (black dots). Plotted is the radial velocity derived from the earlier observation ($v_{\mathrm{rad,early}}$) versus the one derived from the later observation ($v_{\mathrm{rad,late}}$). The inset in the upper left corner shows the comparison for cluster GCs/UCDs. The solid line is the one-to-one relation, the dotted (dashed) line is the fitted relation with slope one for the GC (UCD) data points.}
	\label{fig:veldiff}
\end{figure}

\section{Data reduction and radial velocity measurements}
\label{sec:datared}
For the data reduction from a 2D raw spectrum to a wavelength calibrated 1D spectrum we used the VIPGI (VIMOS Interactive Pipeline and Graphical Interface) data reduction pipeline \citep{2005PASP..117.1284S}. This pipeline is based on the core reduction routines which are also used by the ESO VIMOS pipeline\footnote{\texttt{http://www.eso.org/sci/data-processing/software\linebreak /pipelines/}}. VIPGI is particularly suited for reducing and combining sequences of exposures, and allows amongst others the interactive control of the wavelength calibration of each spectrum.

Since the individual science exposures were taken in at least two different nights, the bias subtraction, flat field division, sky subtraction and wavelength calibration were performed separately for each exposure. Afterwards, the pipeline combined these exposures into a single sky subtracted and wavelength calibrated 2D science frame, and extracted the 1D spectra using the Horne extraction algorithm \citep{1986PASP...98..609H}.

Radial velocities were measured by performing a Fourier cross-correlation between the calibrated 1D object spectra and a template spectrum (IRAF-task \texttt{fxcor} in the \texttt{rv} package). We used a synthetic template spectrum which resembles a typical early-type galaxy \citep{1996AJ....111..603Q}. This template has proven to give the best and most reliable cross-correlation results for this type of survey \citep[e.g.][]{2004A&A...418..445M, 2008A&A...486..697M, 2009A&A...498..705M}.

For a radial velocity measurement to be regarded as reliable, we required a cross-correlation confidence value of $R\geq 5.0$ \citep{1979AJ.....84.1511T}. All spectra for which a lower $R$-value was achieved, were re-measured independently by the authors IM, SM and MH, and the $\vrad$ measurement was only accepted if the three independent measurements were in agreement. Redshifts for objects showing emission lines were determined with the IRAF-task \texttt{rvidlines}. In most of these cases the [OII], H$\gamma$, H$\beta$, [OIII] and H$\alpha$ emission lines were used for measuring the redshift. Heliocentric velocity corrections were applied to all measurements. The individual measurement uncertainties are of the order of 25--160~\kms, with a median of 54~\kms.

\begin{figure}
	\resizebox{\hsize}{!}{\includegraphics{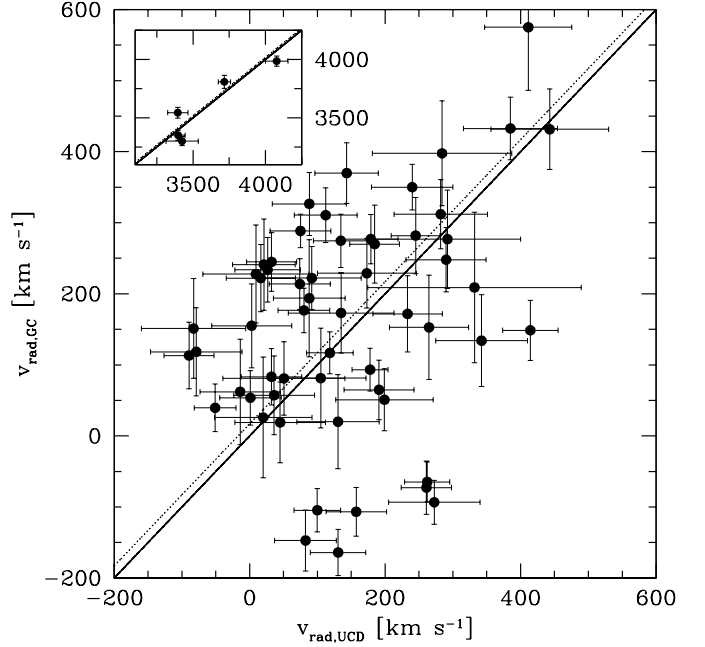}}
	\caption{Comparison of the radial velocities measured in the UCD survey to the radial velocities measured in the GC survey. The dotted line is the fitted relation with slope one. The solid line is the one-to-one relation. The inset in the upper left corner shows the comparison for cluster GCs/UCDs.}
	\label{fig:velcomp}
\end{figure}

\subsection{Systematic radial velocity shifts}
The VIPGI pipeline allows to compute for each slit the offset of a particular sky line in the spectrum from its expected position. With this data it is possible to correct for systematic wavelength shifts which correspond to radial velocity shifts.

Since the wavelength range [5000:7000]~\AA~was used for the radial velocity measurements, we chose the $5577.4$~\AA~[OI] and the $6299.7$~\AA~[OI] sky lines to estimate the systematic wavelength shifts. Given that the shifts were found to be largely independent of the position of the slit on the mask, we recorded the median offset and the according rms of each of the two sky lines for each quadrant and mask. The final radial velocity shift $\Delta \vrad$ was then computed by 
\begin{equation}
 \Delta \vrad = \frac{\delta \lambda}{\lambda_m} \cdot c,
\end{equation}
with the median wavelength offset $\delta \lambda$, the mean wavelength $\lambda_m=5938.6$~\AA, and $c$ the speed of light. The error for $\Delta \vrad$ was calculated from the rms of $\delta \lambda$. The obtained radial velocity shifts were of the order of 40--170~\kms~(or 0.8--3.4~\AA).

Since the observations span a period of several months up to more than one year (see Table~\ref{tab:obslog}), we have to check for systematic radial velocity offsets which are potentially caused by VIMOS instrument instabilities. For this, we compare in Fig.~\ref{fig:veldiff} the measured radial velocities of objects which were observed twice in one individual observing programme, i.e. within the UCD survey (pointing Nr. 1--4), or within the GC survey (pointing Nr. 5--6). For both surveys, a linear fit to the data points gives a relation with a slope consistent with one. Thus, we fix the slope to $m=1$, in order to determine the offset of the fitted relation from the one-to-one relation. For the GC sample, we determine an offset of $\Delta_{\mathrm{GC}} = -17.0 \pm 14.4$~\kms, with a rms of $76.9$~\kms. Within the measurement uncertainties, this is consistent with the one-to-one relation. The clear outlier at $(v_{\mathrm{rad,early}},v_{\mathrm{rad,late}})\sim (4000,3500)$~\kms~was excluded from the fit, since the later measurement was with $R=5.9$ at the border of what we regarded as a reliable measurement (cf. Sect.~\ref{sec:datared}), in contrast to $R=10.3$ for the earlier measurement. 

\begin{figure}
	\resizebox{\hsize}{!}{\includegraphics{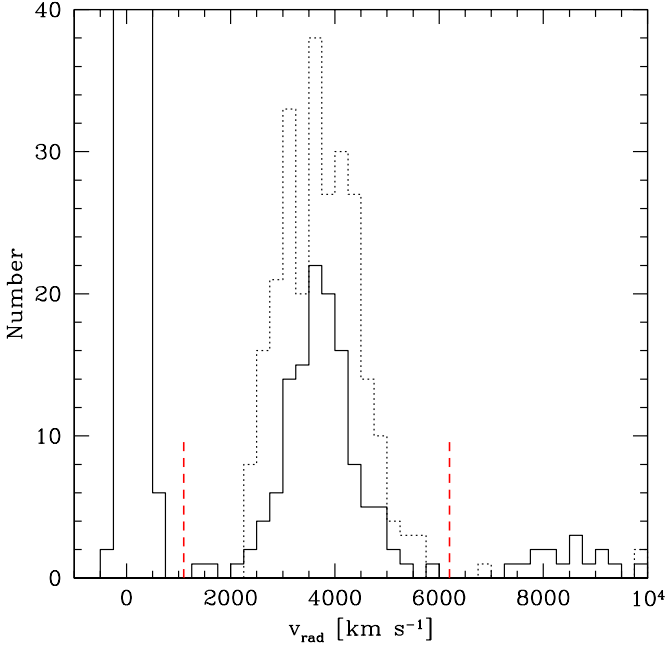}}
	\caption{Radial velocity histogram of objects with $v_{\mathrm{rad}} < 10\,000$~\kms~(solid histogram). The dotted histogram shows the velocity distribution of Hydra\,I cluster galaxies \citep{2003ApJ...591..764C}. The vertical dashed lines mark the velocity limits required for cluster membership in our study.}
	\label{fig:velhist}
\end{figure}

The UCD data points show a larger scatter, and the linear fit results in an offset of $\Delta_{\mathrm{UCD}} = 66.8 \pm 47.9$~\kms, with a rms of $144.8$~\kms. However, there is no clear systematic velocity shift in one direction. For example, data points that compare the earliest with the latest observations lie above as well as below the one-to-one relation. For this reason and due to the rather large measurement uncertainties and the small sample size, we do not apply any constant velocity shift to the UCD data. The larger velocity scatter in the case of the UCD data set may be linked to the commonly known flexure problems of VIMOS, which can lead to misalignment of targets in their respective slits. In fact, many of our UCD targets were located at the very edge of their respective slits, making it in some cases impossible to extract a spectrum.

In Fig.~\ref{fig:velcomp}, the radial velocities measured in the UCD survey are compared to the ones measured in the GC survey. To determine a possible systematic offset, we fit a linear relation with slope $m=1$ to the data. This results in a offset $\Delta_{\mathrm{UCD/GC}} = 17.1 \pm 18.1$~\kms~with a rms of $162.0$~\kms~(dotted line in Fig.~\ref{fig:velcomp}), which is consistent with the one-to-one relation. Again, the relatively large scatter can be attributed to the instrument instabilities of VIMOS.

\begin{figure}
	\resizebox{\hsize}{!}{\includegraphics{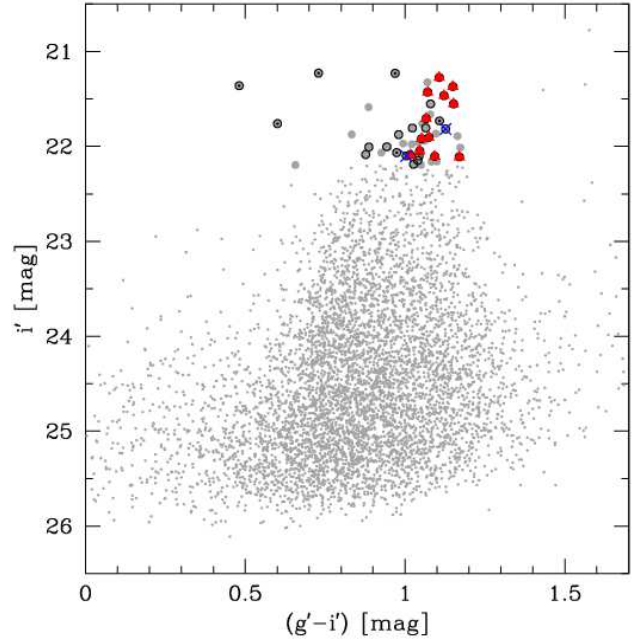}}
	\caption{Colour-magnitude diagram of the globular clusters around NGC~3311 (after W7W8). Note that the magnitudes and colours are not de-reddened. Objects selected for our spectroscopic survey are represented by large grey dots. Open circles mark those objects for which a slit could be allocated. The 12 cluster members are marked by red triangles, the 2 background objects by blue crosses, and the 6 foreground stars by small black dots.}
	\label{fig:wehnercmd}
\end{figure}

\section{Results}
\label{sec:results}
In total, we observed 1236 individual compact objects (including the objects from the previous programme 076.B-0293), compared to 1948 photometrically selected sources (cf. Sect.~\ref{sec:candidates}). Within the surveyed area, the completeness in terms of slit allocation is thus about 63\%. A radial velocity could be measured for 1018 objects ($82$\% of the observed objects). We identified 776 foreground stars, 124 background objects and 118 cluster GCs/UCDs. As the cluster membership criterion we required $1100 < v_{\mathrm{rad}} < 6200$~\kms~(see Fig.~\ref{fig:velhist}). We consider two objects with relatively low radial velocities ($\vrad \sim 1500$~\kms) as cluster members, since also planetary nebulae with similar radial velocities have been identified in the Hydra\,I cluster \citep{2008AN....329.1057V, 2011A&A...528A..24V}. Towards higher radial velocities, there is a clear gap between cluster and background objects, the latter having radial velocities larger than $\vrad \sim 7500$~\kms. A catalogue with the coordinates, magnitudes and colours (obtained from the VIMOS pre-images) and the radial velocities of the 118 cluster objects is given in Table~\ref{tab:hydraucdsgcs} in the appendix (only available on-line).

\begin{figure}
	\resizebox{\hsize}{!}{\includegraphics{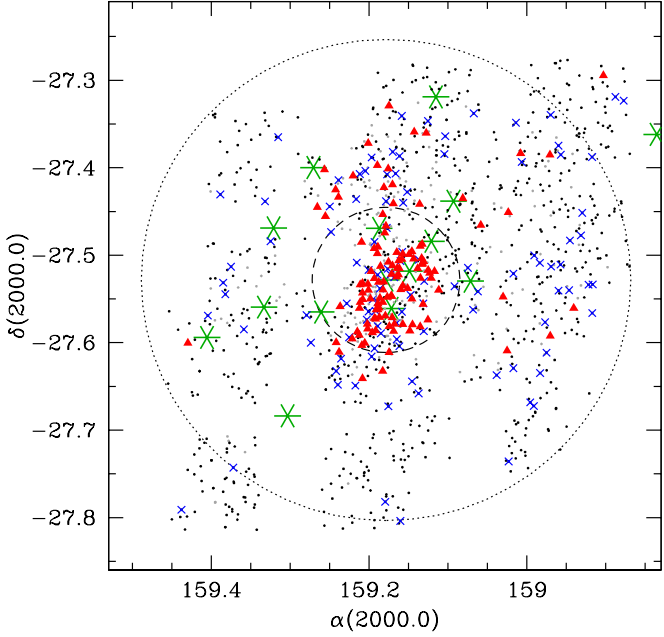}}
	\caption{Spatial distribution of all observed objects. Red triangles are cluster GCs/UCDs, blue crosses mark background objects. Small black dots are foreground stars, and small grey dots are objects for which no radial velocity could be determined. The large green asterisks are major cluster galaxies with apparent magnitudes $R<14$~mag from the spectroscopic study of \citet{2003ApJ...591..764C}. The large dotted circle indicates the projected cluster core-radius of $r_{\mathrm{c}}\sim 170\ h^{-1}$~kpc \citep{1995ApJ...438..527G}, adopting $h=0.75$. The dashed circle marks a radius of $5\arcmin$ around NGC~3311.}
	\label{fig:map}
\end{figure}

For the sub-sample of slightly resolved UCD candidates (see Sect.~\ref{sec:ucdcands}), we selected 35 objects, of which 27 were observed, yielding 20 background objects and 4 foreground stars. No radial velocity could be measured for the remaining 3 objects. 

30 of the 48 selected UCD candidates from W7W8 could be observed. Figure~\ref{fig:wehnercmd} shows a colour-magnitude diagram (CMD) of the globular clusters around NGC~3311, as obtained from their Gemini South GMOS imaging in $g^\prime$ and $i^\prime$. The spectroscopically observed objects are highlighted. For 20 of them we were able to measure a radial velocity, resulting in 12 cluster members, 6 foreground stars and 2 background objects.

In Fig.~\ref{fig:map} we present a map of the entire sample of objects observed in our survey.

\subsection{Photometric and structural properties}

\subsubsection{VLT/VIMOS imaging}
\label{sec:photprops}
The objects' instrumental magnitudes, as measured in the pre-images, were calibrated using the photometric zeropoints given by the ESO Quality Control and Data Processing Group\footnote{QCG, \texttt{http://www.eso.org/observing/dfo/quality/}}. Colours and magnitudes were then corrected for interstellar absorption and reddening, $A_V=0.26$~mag and $E(V-R)=0.05$~mag \citep{1998ApJ...500..525S}.

\begin{figure}
	\resizebox{\hsize}{!}{\includegraphics{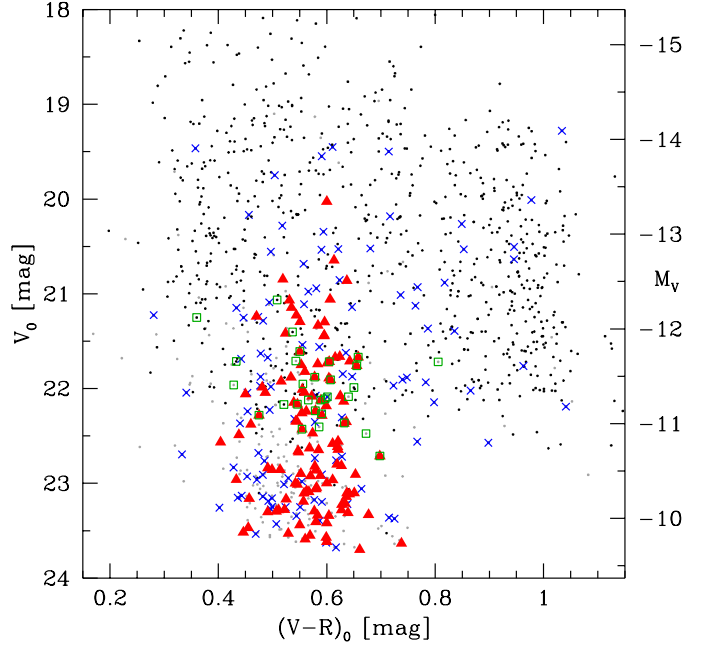}}
	\caption{Colour-magnitude diagram of the objects from Fig.~\ref{fig:map}, except for the major cluster galaxies. Green open squares mark the observed UCD candidates from W7W8. The right axis indicates the absolute $V$-band magnitude $M_V$.}
	\label{fig:cmd}
\end{figure}

A CMD of all observed objects is plotted in Fig.~\ref{fig:cmd}. Adopting a Hydra\,I distance modulus of $(m-M)=33.37$~mag \citep{2005A&A...438..103M}, 52 of the identified cluster objects are brighter than $M_V = -11.0$~mag (see also Table~\ref{tab:hydraucdsgcs}). This luminosity corresponds to a mass of $\gtrsim 6.5 \times 10^6$~solar masses, applying a mean UCD $M/L$ ratio of 3 \citep[e.g.][]{2008A&A...487..921M}. Such objects are generally called UCDs, although the distinction between UCDs and GCs is not clear at low luminosities. Even if assuming $M/L=1$, these objects are still more luminous and more massive than $\omega$~Centauri, which is with $M_V=-10.29$~mag and a mass of $2.5\times 10^6$~M$_{\sun}$ the most luminous and most massive globular cluster in our Galaxy \citep{1996AJ....112.1487H,2006A&A...445..513V}. Adopting this mass limit for the separation of GCs and UCDs, as suggested by \citet{2005ApJ...627..203H} and \citet{2008A&A...487..921M}, we even have more than 80 UCDs in our sample. In comparison to the photometric studies of W7W8, we find 15 UCDs brighter than $M_V = -11.8$~mag, which is the magnitude of the brightest confirmed cluster object from their list of UCD candidates (see Fig.~\ref{fig:cmd}). The three brighter candidate objects turned out to be foreground stars.

\begin{figure}
	\resizebox{\hsize}{!}{\includegraphics{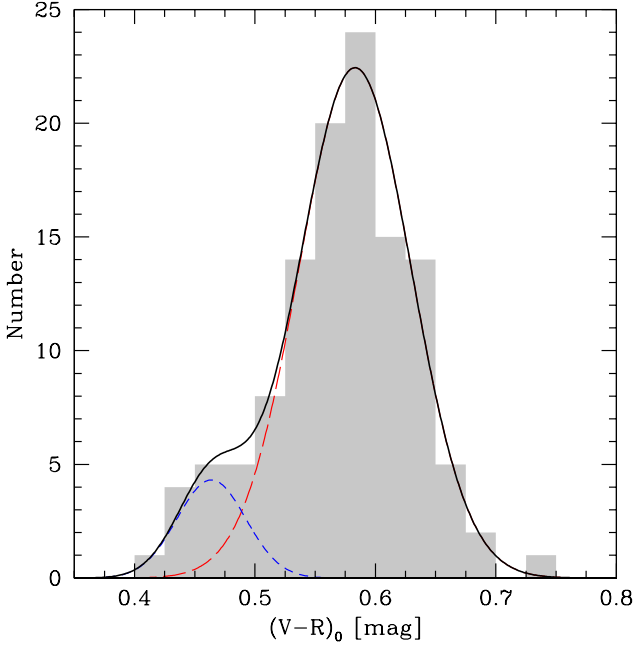}}
	\caption{Colour distribution of confirmed Hydra\,I GCs/UCDs (shaded histogram). The black curve is a double peak Gaussian function fitted to the data, the dashed curves are the single components.}
	\label{fig:colorhist}
\end{figure}

The identified cluster GCs/UCDs have de-reddened colours of $0.40 < (V-R)_0 < 0.75$~mag, but they are not uniformly distributed in colour-space, as Fig.~\ref{fig:colorhist} shows. A bimodal GC colour distribution is expected for NGC~3311 \citep{2008ApJ...681.1233W}, and for our data, a double Gaussian distribution is preferred to a single Gaussian distribution ($\chi^2_\nu = 1.141$ and $\chi^2_\nu =2.784$, respectively). Fitting a double Gaussian function to the data results in a blue peak at $(V-R)_0 = 0.46$~mag and a red peak at $(V-R)_0 = 0.58$~mag. For a 13-Gyr population, these peaks correspond to metallicities of $\mathrm{[Fe/H]} \approx -2.2$~dex and $\mathrm{[Fe/H]} \approx -0.4$~dex, respectively \citep{2003MNRAS.344.1000B}. These estimates are consistent with the values derived for metal-poor and metal-rich GCs in a number of giant elliptical galaxies \citep[][and references therein]{2006ARA&A..44..193B}. Given this, we define for the following analyses (Sect.~\ref{sec:kinematics}) a blue (metal-poor) and a red (metal-rich) sub-population with $(V-R)_0 < 0.5$~mag and $(V-R)_0 \geq 0.5$~mag, respectively. The separating colour corresponds to a metallicity of $\mathrm{[Fe/H]} \approx -1.4$~dex.

With $(V-R)_0 = 0.60$~mag, the brightest UCD in our sample (HUCD1) clearly belongs to the metal-rich sub-population and is located at the tip of the red GC sequence in the CMD. This confirms results of photometric studies, where the red GC sequence is found to extend to higher luminosities than the blue sequence \citep[e.g.][]{2008ApJ...681.1233W, 2010ApJ...710.1672M}.

\subsubsection{HST/WFPC2 imaging}
\label{sec:hstimaging}
For 26 of the identified cluster GCs/UCDs, imaging is available in the HST archive. Two WFPC2 fields were observed in Cycle~6 with the PC1 chip centred on NGC~3311 and NGC~3309, respectively. The exposure times were $3700$~s in $F555W$ and $3800$~s in $F814W$ for the field centred on NGC~3311, and $4400$~s in both filters for the NGC~3309 field (HST programme GO.06554.01-95A, PI: J.P. Brodie, see also \citealt{2000ApJ...543L..19B}).

The F555W and F814W images were bias, flat-field and bad pixel corrected. The sub-pixel dithered imaging allowed us to eliminate cosmic-rays, hot pixels and improve the spatial resolution. To perform a PSF photometry on each individual WFPC2 detector, we created a grid of $100\times100$~PSFs with the TinyTim\footnote{TinyTim accounts for significant variation of the PSF as a function of wavelength (filter) due to diffraction, large angle scattering, field-dependence, aberrations, focus offsets between cameras and wavelength dependent charge diffusion (see \texttt{http://www.stsci.edu/software/tinytim/})} software package \citep{1995ASPC...77..349K}. This library was used to create a spatially variable PSF model and perform PSF-fitting photometry with the \texttt{allstar} task in IRAF. Isolated stars were used to determine an aperture correction to $1 \farcs 0$ aperture diameter, which is the aperture used by \citet{2009PASP..121..655D} to derive the most up to date CTE corrections and photometric zero points for each of the WFPC2 chips. Finally, the VEGAMAG WFPC2 photometry was corrected for foreground extinction using \citet{1998ApJ...500..525S} dust maps towards the direction of both galaxies, and \citet{1989ApJ...345..245C} relations for $R_V=3.1$ to calculate the absorption at the effective wavelengths for the WFPC2 filters. The following dereddening values were applied: $A_{F555W}=0.261$~mag and $A_{F814W}=0.156$~mag. To transform the VEGAMAG WFPC2 magnitudes to the standard Johnson/Cousins magnitudes, we used the \citet{2009PASP..121..655D} transformations with the coefficients in their table~4 for the $(V-I)$ colour.

\begin{figure}
	\resizebox{\hsize}{!}{\includegraphics{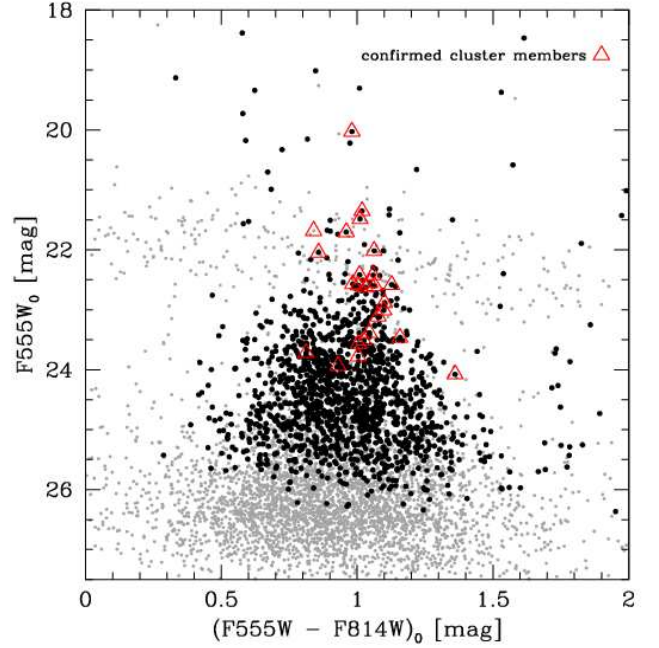}}
	\caption{Colour-magnitude diagram of all WFPC2 sources around NGC~3311 and NGC~3309. Black dots are objects with colour errors $(F555W - F814W)_0 < 0.1$~mag, grey dots are those with larger colour errors. Red open triangles mark the spectroscopically confirmed cluster GCs/UCDs.}
	\label{fig:hstcmd}
\end{figure}

\begin{table*}
\renewcommand{\arraystretch}{1.2}
	\caption{Photometric and structural properties of cluster GCs/UCDs with available HST imaging. Listed are the extinction corrected apparent $F555W_0$, $F814W_0$, and $V_0$ magnitudes in columns 2--4 and the $(V-I)_0$ colour in the fifth column. Columns~6 and 7 give the half-light radius $r_{\mathrm{h}}$ in pixels and in pc, respectively. Column~8 lists the tidal-to-core radius $r_{\mathrm{t}}/r_{\mathrm{c}}$ of the best fit model, and the last column gives the signal-to-noise ratio of the source detection.}
	\label{tab:hstucds}
	\centering	
		\begin{tabular}{r c c c c r r c r}
		\hline\hline
		ID & $F555W_0$ & $F814W_0$ & $V_0$ & $(V-I)_0$ & $r_{\mathrm{h}}$ & $r_{\mathrm{h}}$ & $r_{\mathrm{t}}/r_{\mathrm{c}}$ & $S/N$ \\
		   & [mag] & [mag] & [mag] & [mag] & [pix] & [pc] & &  \\
		\hline
HUCD1 & $ 20.027 \pm  0.076 $ & $ 19.046 \pm 0.085 $ & $ 20.00 \pm  0.08 $ & $  0.99 \pm  0.11 $ & $  1.11^{+ 0.01}_{-0.01} $ & $  25.4^{+  0.2}_{ -0.2} $ & 5 & 134.4 \\ 
HUCD8 & $ 21.350 \pm  0.053 $ & $ 20.332 \pm 0.072 $ & $ 21.32 \pm  0.02 $ & $  1.03 \pm  0.03 $ & $  0.81^{+ 0.03}_{-0.02} $ & $  18.5^{+  0.7}_{ -0.5} $ & 15 &  62.6 \\ 
HUCD14 & $ 21.483 \pm  0.024 $ & $ 20.472 \pm 0.032 $ & $ 21.46 \pm  0.02 $ & $  1.02 \pm  0.04 $ & $  0.63^{+ 0.01}_{-0.02} $ & $  25.6^{+  0.9}_{ -1.1} $ & 5 &  64.0 \\ 
HUCD18 & $ 22.044 \pm  0.026 $ & $ 21.186 \pm 0.011 $ & $ 22.02 \pm  0.03 $ & $  0.87 \pm  0.03 $ & $  0.91^{+ 0.05}_{-0.09} $ & $  13.5^{+  1.8}_{ -1.6} $ & 30 &  55.1 \\ 
HUCD21 & $ 21.703 \pm  0.022 $ & $ 20.743 \pm 0.030 $ & $ 21.68 \pm  0.02 $ & $  0.97 \pm  0.04 $ & $  0.52^{+ 0.03}_{-0.02} $ & $  11.9^{+  0.7}_{ -0.5} $ & 5 &  50.1 \\ 
HUCD30 & $ 22.016 \pm  0.024 $ & $ 20.953 \pm 0.039 $ & $ 21.99 \pm  0.02 $ & $  1.08 \pm  0.05 $ & $  0.57^{+ 0.03}_{-0.03} $ & $  13.1^{+  0.7}_{ -0.7} $ & 15 &  47.8 \\ 
HUCD35 & $ 21.681 \pm  0.088 $ & $ 20.841 \pm 0.084 $ & $ 21.66 \pm  0.03 $ & $  0.85 \pm  0.04 $ & $  1.12^{+ 0.04}_{-0.05} $ & $  10.1^{+  0.7}_{ -0.9} $ & 5 &  43.9 \\ 
HUCD39 & $ 22.569 \pm  0.019 $ & $ 21.585 \pm 0.033 $ & $ 22.54 \pm  0.02 $ & $  1.00 \pm  0.04 $ & $  0.37^{+ 0.05}_{-0.04} $ & $   8.5^{+  1.1}_{ -0.9} $ & 30 &  29.4 \\ 
HUCD42 & $ 22.424 \pm  0.029 $ & $ 21.365 \pm 0.032 $ & $ 22.40 \pm  0.03 $ & $  1.07 \pm  0.04 $ & $  0.49^{+ 0.05}_{-0.02} $ & $   9.8^{+  1.1}_{ -0.9} $ & 5 &  37.2 \\ 
HUCD44 & $ 22.416 \pm  0.024 $ & $ 21.407 \pm 0.023 $ & $ 22.39 \pm  0.02 $ & $  1.02 \pm  0.03 $ & $  0.41^{+ 0.04}_{-0.04} $ & $   9.4^{+  0.9}_{ -0.9} $ & 30 &  37.3 \\ 
HUCD46 & $ 22.625 \pm  0.018 $ & $ 21.594 \pm 0.029 $ & $ 22.60 \pm  0.02 $ & $  1.04 \pm  0.03 $ & $  0.44^{+ 0.03}_{-0.04} $ & $  14.0^{+  1.1}_{ -1.4} $ & 15 &  28.2 \\ 
HUCD47 & $ 22.590 \pm  0.024 $ & $ 21.586 \pm 0.045 $ & $ 22.56 \pm  0.02 $ & $  1.02 \pm  0.05 $ & $  0.38^{+ 0.03}_{-0.03} $ & $   8.7^{+  0.7}_{ -0.7} $ & 30 &  34.0 \\ 
54 & $ 22.550 \pm  0.040 $ & $ 21.503 \pm 0.049 $ & $ 22.52 \pm  0.04 $ & $  1.06 \pm  0.06 $ & $  0.46^{+ 0.04}_{-0.05} $ & $  10.5^{+  0.9}_{ -1.1} $ & 30 &  27.9 \\ 
59 & $ 22.601 \pm  0.027 $ & $ 21.536 \pm 0.032 $ & $ 22.58 \pm  0.03 $ & $  1.08 \pm  0.04 $ & $  0.58^{+ 0.04}_{-0.03} $ & $  11.0^{+  0.7}_{ -0.7} $ & 5 &  27.0 \\ 
61 & $ 23.108 \pm  0.029 $ & $ 22.031 \pm 0.046 $ & $ 23.08 \pm  0.03 $ & $  1.09 \pm  0.05 $ & $  0.47^{+ 0.05}_{-0.07} $ & $  10.8^{+  1.1}_{ -1.6} $ & 30 &  22.6 \\ 
66 & $ 22.582 \pm  0.035 $ & $ 21.455 \pm 0.035 $ & $ 22.56 \pm  0.03 $ & $  1.14 \pm  0.05 $ & $  0.50^{+ 0.05}_{-0.03} $ & $  11.5^{+  1.1}_{ -0.7} $ & 5 &  27.4 \\ 
68 & $ 22.880 \pm  0.029 $ & $ 21.779 \pm 0.037 $ & $ 22.86 \pm  0.03 $ & $  1.12 \pm  0.05 $ & $  0.50^{+ 0.05}_{-0.03} $ & $   9.6^{+  0.5}_{ -0.9} $ & 5 &  21.4 \\ 
69 & $ 23.009 \pm  0.031 $ & $ 21.911 \pm 0.024 $ & $ 22.98 \pm  0.03 $ & $  1.11 \pm  0.04 $ & $  0.49^{+ 0.04}_{-0.05} $ & $  11.2^{+  0.9}_{ -1.1} $ & 5 &  20.8 \\ 
90 & $ 23.511 \pm  0.033 $ & $ 22.486 \pm 0.045 $ & $ 23.49 \pm  0.03 $ & $  1.04 \pm  0.06 $ & $  0.61^{+ 0.05}_{-0.06} $ & $   9.6^{+  1.4}_{ -1.6} $ & 5 &  14.8 \\ 
94 & $ 23.463 \pm  0.029 $ & $ 22.305 \pm 0.016 $ & $ 23.44 \pm  0.03 $ & $  1.17 \pm  0.03 $ & $  0.42^{+ 0.06}_{-0.07} $ & $  11.5^{+  2.1}_{ -1.8} $ & 15 &  17.2 \\ 
99 & $ 23.385 \pm  0.020 $ & $ 22.340 \pm 0.036 $ & $ 23.36 \pm  0.02 $ & $  1.06 \pm  0.04 $ & $  0.50^{+ 0.05}_{-0.05} $ & $  11.5^{+  1.1}_{ -1.1} $ & 5 &  17.4 \\ 
101 & $ 23.570 \pm  0.026 $ & $ 22.565 \pm 0.038 $ & $ 23.54 \pm  0.03 $ & $  1.02 \pm  0.05 $ & $  0.43^{+ 0.05}_{-0.04} $ & $  13.3^{+  0.9}_{ -0.7} $ & 5 &  13.1 \\ 
110 & $ 23.722 \pm  0.045 $ & $ 22.910 \pm 0.035 $ & $ 23.70 \pm  0.04 $ & $  0.82 \pm  0.06 $ & $  0.56^{+ 0.09}_{-0.06} $ & $  12.8^{+  2.1}_{ -1.4} $ & 5 &  12.1 \\ 
112 & $ 23.935 \pm  0.039 $ & $ 23.005 \pm 0.053 $ & $ 23.91 \pm  0.04 $ & $  0.94 \pm  0.07 $ & $  0.40^{+ 0.10}_{-0.07} $ & $   9.2^{+  2.3}_{ -1.6} $ & 15 &  10.6 \\ 
114 & $ 23.775 \pm  0.027 $ & $ 22.771 \pm 0.033 $ & $ 23.75 \pm  0.03 $ & $  1.02 \pm  0.04 $ & $  0.59^{+ 0.08}_{-0.07} $ & $  14.4^{+  0.2}_{ -0.5} $ & 5 &  11.7 \\ 
115 & $ 24.077 \pm  0.036 $ & $ 22.716 \pm 0.035 $ & $ 24.06 \pm  0.04 $ & $  1.38 \pm  0.05 $ & $  0.50^{+ 0.09}_{-0.08} $ & $  11.2^{+  1.1}_{ -0.5} $ & 5 &  10.5 \\ 
		\hline
		\end{tabular}
\end{table*}

At a the adopted distance of $47.2$~Mpc ($m-M = 33.37$~mag), one PC1 and WF2,3,4 pixel corresponds to 11 and 22~pc, respectively. For high signal-to-noise objects (typically $S/N \gtrsim 30$), one can reliably measure $r_{\mathrm{h}}$ down to 0.1~pix, i.e. $\sim 2$~pc. Therefore, it is feasible to measure sizes for extended objects (such as UCDs and bright GCs) from the WFPC2 frames.

To measure the half-light radii $r_{\mathrm{h}}$ of the 26 spectroscopically confirmed objects, we generated ten times sub-sampled PSFs with TinyTim for $F555W$. Each PSF was tailored to the position of the object on the chip. Utilizing the \texttt{ishape} task of the \texttt{baolab} software package\footnote{\texttt{http://www.astro.uu.nl/\~{}larsen/baolab/}} \citep{1999A&AS..139..393L}, this PSF was used to model the object profile as an analytical function convolved with the (model) PSF. When sub-sampling is enabled, TinyTim does not include a convolution with the charge diffusion kernel (CDK), which additionally smears the stellar PSF. Thus, during fitting with \texttt{ishape}, the TinyTim PSF was convolved with a $F555W$ CDK, that simulates blurring caused by charge diffusion which is well understood for the $F555W$ filter.

All objects were modelled with \citet{1962AJ.....67..471K} profiles with concentrations of the tidal-to-core radius of $r_{\mathrm{t}}/r_{\mathrm{c}}=5, 15, 30$ and 100. We adopted the structural parameter measurements from the best $\chi^2$ fit model. The output $r_{\mathrm{h}}$ is the $r_{\mathrm{h}}$ along the semi-major axis which needs to be corrected for ellipticity and brought to the geometrical mean value ("effective" $r_{\mathrm{h}}$) by multiplying the square root of the major/minor axis ratio \citep[for details see e.g. Eq.~1 in][]{2008AJ....135.1858G}.

Table~\ref{tab:hstucds} lists the photometric and structural properties of the 26 cluster GCs/UCDs. The $V_0$ magnitudes derived from the HST images are, on average, 0.2--0.4~mag fainter than those measured in the VIMOS pre-images. Reasons for this discrepancy likely include the uncertainty in the VIMOS photometric zeropoints (no photometric standard was taken at the night of the pre-imaging), the different magnitude measurement techniques, and the uncertainty in the magnitude transformation of the HST data.

\begin{figure}
	\resizebox{\hsize}{!}{\includegraphics{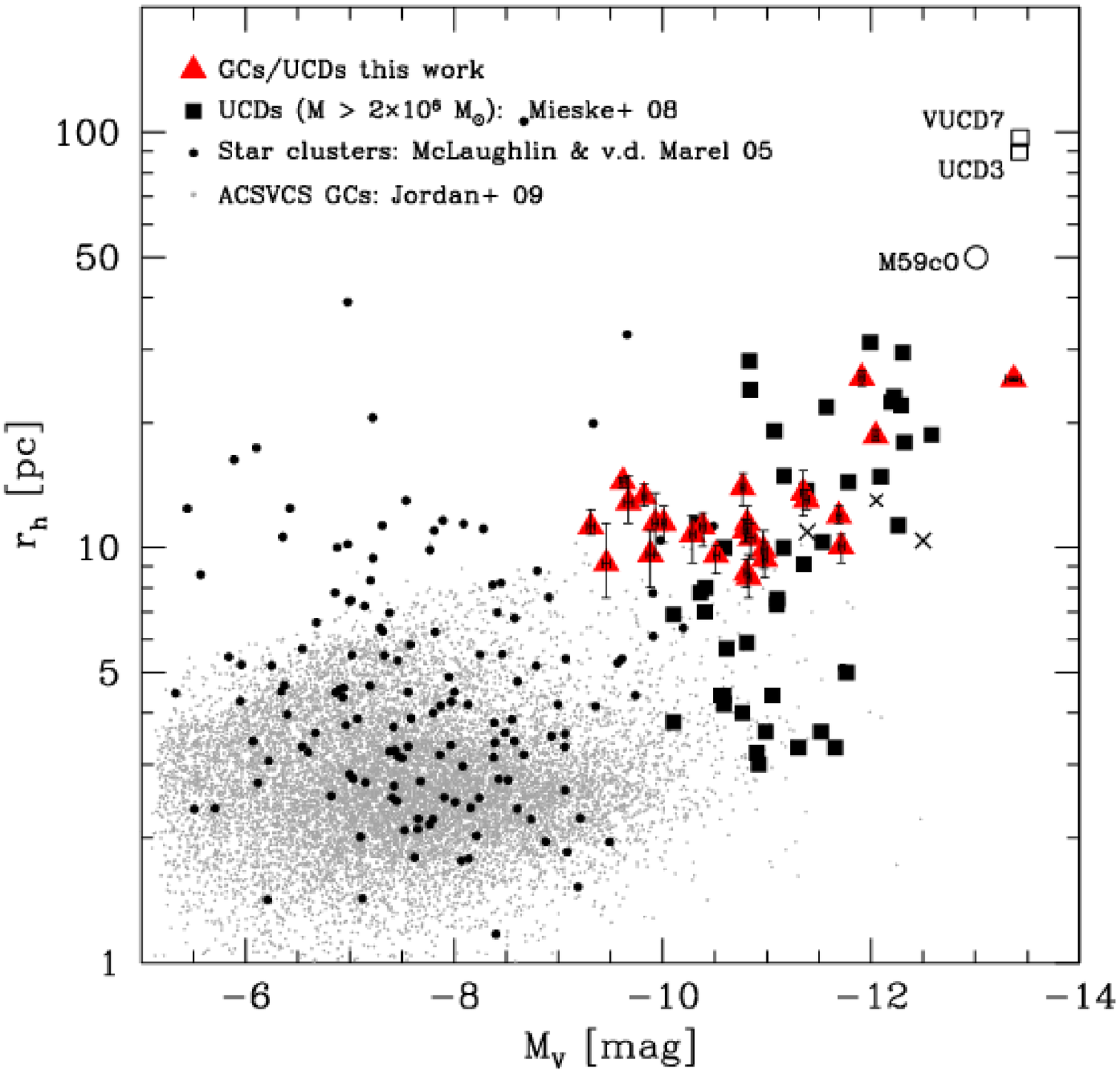}}
	\caption{Half-light radii and luminosities of the 26 cluster GCs/UCDs with HST imaging (see Table~\ref{tab:hstucds}), in comparison to other star clusters and UCDs. Small crosses indicate the \textit{core} components of VUCD7, UCD3 \citep[both from][]{2007AJ....133.1722E}, and M59cO \citep{2008MNRAS.385L..83C}.}
	\label{fig:hstucds}
\end{figure}

Figure~\ref{fig:hstcmd} shows a colour-magnitude diagram of all WFPC2 sources around NGC~3309/3311, with the confirmed cluster GCs/UCDs highlighted. In Fig.~\ref{fig:hstucds}, we compare the half-light radii and luminosities of the Hydra\,I GCs/UCDs to globular clusters from the ACS Virgo Cluster Survey \citep[ACSVCS,][]{2009ApJS..180...54J}, Milky Way, LMC/SMC and Fornax star clusters from \citet{2005ApJS..161..304M}, UCDs from \citet{2008A&A...487..921M}, and the compact object M59cO \citep{2008MNRAS.385L..83C}. 

The apparent $g_{\mathrm{AB}}$-band magnitudes of the ACSVCS GCs were transformed into absolute $V$-band magnitudes using the relation $V = g_{\mathrm{AB}} + 0.026-0.307 \cdot (g-z)_{\mathrm{AB}}$ given in \citet{2006ApJ...639..838P}, and a Virgo distance modulus of $31.09$~mag \citep{2007ApJ...655..144M}. $r_{\mathrm{h}}$ is the average of the half-light radii measured in the $g$- and in the $z$-band. $M_V$ and $r_{\mathrm{h}}$ of the star clusters from \citet{2005ApJS..161..304M} are the King models values. The tabulated masses of the UCDs from \citet{2008A&A...487..921M} were converted into $M_V$ with the given $M/L_V$ ratios and a solar absolute magnitude of $M_{V,\sun} = 4.83$~mag \citep{1998gaas.book.....B}. For M59cO, $M_V$ was calculated from $M_B$ using $B-V=0.96$~mag \citep{1995PASP..107..945F}.

\begin{figure}
	\resizebox{\hsize}{!}{\includegraphics{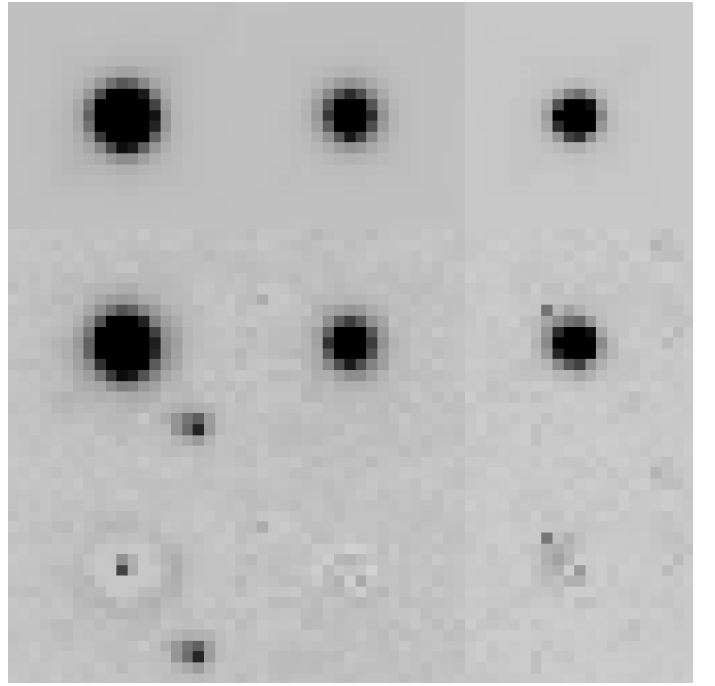}}
	\caption{Residual images for the three brightest UCDs with HST imaging. From left to right are sorted by luminosity: HUCD1, HUCD8 and HUCD14. From top to bottom are shown the PSF model, the object and the residual image. Each panel is $20\times20$~pixels in size.}
	\label{fig:hstimages}
\end{figure}

The sizes, luminosities and colours of the Hydra\,I GC/UCDs are fully consistent with the ones of Virgo and Fornax UCDs \citep[e.g.][]{2008AJ....136..461E, 2008A&A...487..921M}. A size-luminosity relation is visible for objects brighter than $M_V \sim -10$~mag, following the trend observed in other galaxy clusters. Below this magnitude, we do not regard the size measurements with \texttt{ishape} reliable, since the $S/N$ is smaller than 20 for those objects (cf. Table~\ref{tab:hstucds}).

With $M_V=-13.37$~mag, HUCD1 is the brightest object in our sample. Its luminosity corresponds to a mass of about $5\times 10^7$~M$_{\sun}$ (assuming $M/L=3$), or even $\sim 10^8$~M$_{\sun}$, if assuming a mass-to-light ratio larger than 5, which has been measured for several of the brightest UCDs in Virgo and Fornax \citep{2008A&A...487..921M}. Hence, HUCD1 has a luminosity/mass comparable to the most massive UCDs in Virgo and Fornax (VUCD7 and UCD3), but with its half-light radius of $r_{\mathrm{h}}=25.4$~pc, it is the most compact object among the highest luminosity UCDs. VUCD7 and UCD3 are known to feature a two-component light profile with an extended faint envelope and a smaller core component \citep{2007AJ....133.1722E}, which is also indicated in Fig.~\ref{fig:hstucds}. A weak indication for a faint halo is observed for HUCD1, but not for the other two luminous UCDs. This can be seen in Fig.~\ref{fig:hstimages}, where from from left to right are shown the three brightest UCDs in our sample, and from top to bottom the according PSF model, the object and the residual (object minus PSF model) image. The residuals for HUCD1 are of the order of a few percent of the value of the corresponding science image pixel. However, King profiles are known not to represent well the outer regions of GCs \citep[e.g.][]{2005ApJS..161..304M}, therefore we cannot affirmatively conclude that the observed residual halo of HUCD1 is due to a presence of a second component as for VUCD7 and UCD3.

\begin{figure}
	\resizebox{\hsize}{!}{\includegraphics{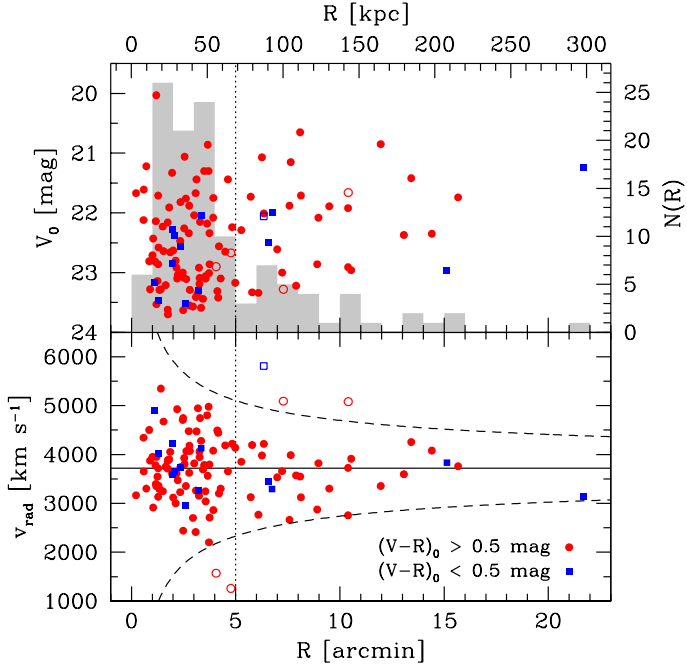}}
	\caption{Apparent magnitude $V_0$ and radial velocity $v_{\mathrm{rad}}$ of confirmed cluster GCs/UCDs as a function of $R$, the projected distance from NGC~3311. The histogram in the \textit{upper panel} shows the number counts in radial bins of $1\arcmin$ width. The dashed curves in the \textit{lower panel} envelope the objects that remain after applying the rejection algorithm (see text for details). They are of the form $v_{\mathrm{env}}(R)=v_{\mathrm{sys}} \pm \sqrt{C_{\mathrm{max}} / R}$, where $C_{\mathrm{max}}$ is the product $v^2\cdot R$ for the first object that is not rejected, and $v_{\mathrm{sys}}$ is the mean radial velocity of all objects (solid line). Open symbols denote the rejected objects. The vertical dotted line divides the inner from the outer sample.}
	\label{fig:radialdist}
\end{figure}

\subsection{Kinematics}
\label{sec:kinematics}
The majority of the confirmed cluster GCs/UCDs is located in the immediate vicinity of NGC~3311, i.e. within a projected distance of $R=5\arcmin$, or $\sim 70$~kpc (see Fig.~\ref{fig:map}). In Fig.~\ref{fig:radialdist} we plot the objects apparent magnitudes and radial velocities versus the projected distance from NGC~3311. There appears to be a trend that brighter objects are in projection located further away from the central galaxy. This is, however, caused by selection effects, in the sense that the fields in which brighter objects have been selected reach further out in radius (cf. Fig.~\ref{fig:fields} and Sect.~\ref{sec:ucdcands}).

In order to clean the sample, we applied an outlier rejection method as described in \citet{2010A&A...513A..52S} to our data. This method is based on the tracer mass estimator by \citet{2003ApJ...583..752E}. In a first step, the quantity
\begin{equation}
\label{eq:mn}
m_N=\frac{1}{N}\sum_{i=1}^{N} {v_i}^2\cdot R_i
\end{equation}
is calculated,  where $v_i$ are the velocities of the considered objects, relative to the mean velocity $v_{\mathrm{sys}}=3717$~\kms~of the entire sample. $R_i$ are the projected distances from NGC~3311, and $N$ is the number of objects. Then, in an iterative process, the object with the largest contribution to $m_N$, i.e. $\max(v^2\cdot R)$, is removed and $m_N$ is again calculated for the remaining $N-1$ objects. In this way, 5 outliers were identified (open symbols in Fig.~\ref{fig:radialdist}).

The mean radial velocity $\bar{v}_{\mathrm{rad}}$ and the dispersion $\sigma$ of the full and the cleaned sample were determined with the maximum likelihood estimator function \texttt{fitdistr}, which is implemented in the \texttt{R}-statistics software\footnote{\citet{Rstatistics}, \texttt{http://www.r-project.org}.}. The values for $\bar{v}_{\mathrm{rad}}$ and $\sigma$ before and after applying the outlier rejection method are listed in Table~\ref{tab:velcolhist}. The table also gives the dispersion $\sigma_{\mathrm{PM}}$, as returned by the \citet{1993ASPC...50..357P} estimator. This estimator additionally takes into account the individual measurement errors, i.e. the velocities are weighted by their respective uncertainties.

\subsubsection{Mean radial velocities}
\label{sec:meanvelocities}
\begin{figure}
	\resizebox{\hsize}{!}{\includegraphics{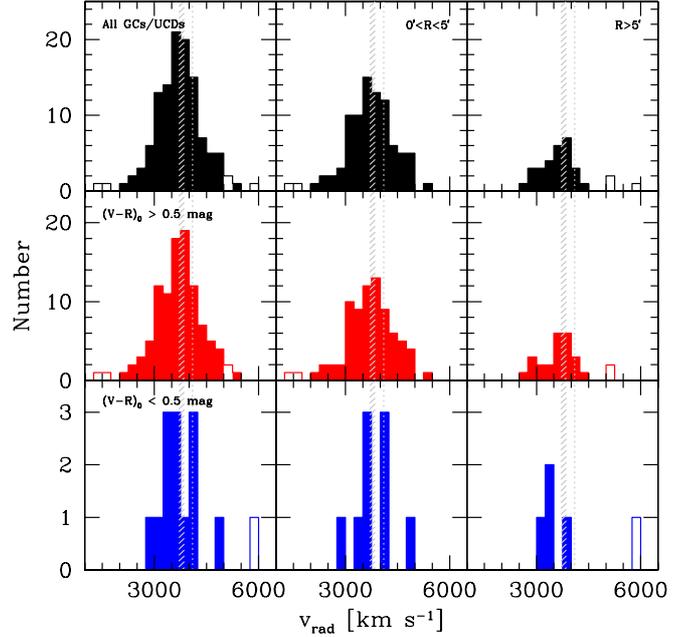}}
	\caption{Radial velocity distribution of all identified GCs/UCDs (\textit{top left panel}), the red sub-population (\textit{middle left panel}), and the blue sub-population (\textit{bottom left panel}). Each population is radially binned in the panels of the \textit{middle column} ($0'<R<5'$) and the \textit{right column} ($R>5'$). In all panels, the open histogram represents the rejected objects, according to Fig.~\ref{fig:radialdist}. The grey dashed area marks the range of radial velocities reported for NGC~3311 in the literature. The dotted vertical line indicates the systemic velocity of NGC~3309 at $\sim$4100~\kms.}
	\label{fig:velcolhist}
\end{figure}

The radial velocity distributions of the entire GC/UCD population, the red and the blue sub-population (see Sect.~\ref{sec:photprops} and Fig.~\ref{fig:colorhist}) are shown in Fig.~\ref{fig:velcolhist}. Each population is also sub-divided into a central population ($0'<R<5'$) and an outer population ($R>5'$). The mean radial velocities of the different GC/UCD samples are consistent with the range of radial velocities reported for NGC~3311 in the literature, i.e. $3700 \lesssim \vrad \lesssim 3850$~\kms~\citep[e.g.][]{1995ApJ...440...28P, 2003ApJ...591..764C, 2003AJ....126.2268W, 2008A&A...486..697M}. This suggests that the GC/UCD system is dynamically rather associated to the cD galaxy NGC~3311, instead of belonging to the close-by giant elliptical galaxy NGC~3309, which has a significantly higher radial velocity of $\sim$4100~\kms~\citep[e.g.][]{2003ApJ...591..764C, 2003AJ....126.2268W, 2008A&A...486..697M}. This result is supported by a detailed photometric study of the globular cluster system around NGC~3311 and NGC~3309 \citep{2008ApJ...681.1233W}. The authors show that the GC system of NGC~3311 is completely dominant in terms of specific frequency $S_N$.

Figures~\ref{fig:map} and \ref{fig:radialdist} show that the outer GCs/UCDs are spaciously distributed between 70 and 200~kpc away from the cluster centre. Only three of these objects are in projection located close to a major cluster galaxy. The full sample of 31 objects has a mean radial velocity which is consistent with the systemic velocity of NGC~3311, while the outlier-cleaned sample (28 objects) has a $\sim 170$~\kms~lower mean radial velocity. However, both values agree within their uncertainties. These results suggest that most of the outer objects also belong to the NGC~3311 GC system. This is consistent with studies of the GC system of NGC~1399, the central galaxy of the Fornax cluster, where its GC system could be traced out to a distance of $\sim 250$~kpc \citep{2006A&A...451..789B, 2008A&A...477L...9S}.

\begin{figure*}
	\resizebox{\hsize}{!}{\includegraphics{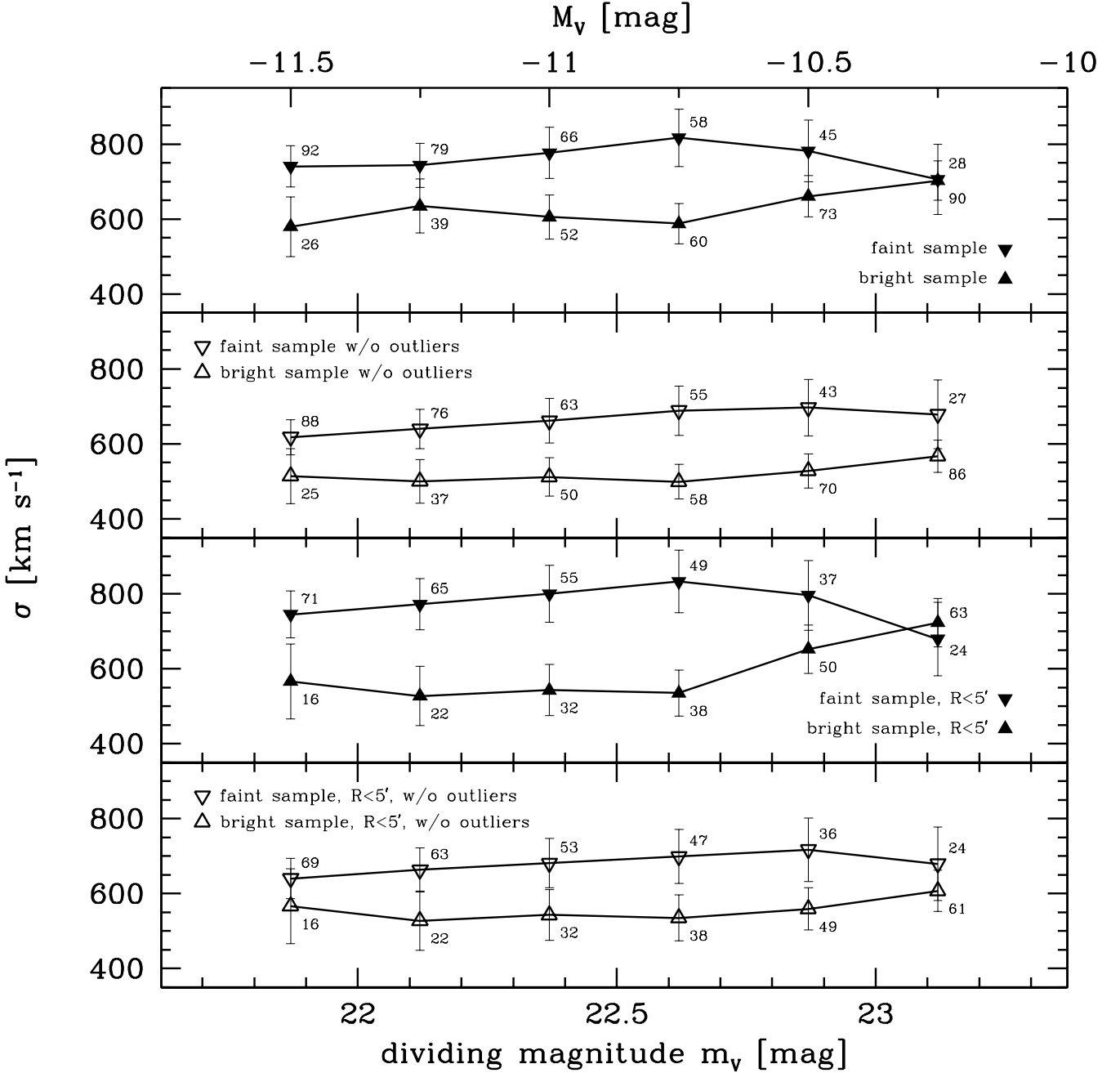}}
	\caption{Velocity dispersion $\sigma$ of a bright and a faint sample as a function of the dividing magnitude $m_V$. The upper two panels show objects in the full radial range, the lower two panels only show objects with $R<5\arcmin$ (cf. Fig.~\ref{fig:radialdist}). The number of objects contained in the bright/faint sample is indicated next to each data point. The comparison is done for both the full sample (filled symbols) and the outlier-cleaned sample (open symbols).}
	\label{fig:magdisp}
\end{figure*}

\begin{table}
	\caption{Mean radial velocities $\bar{v}_{\mathrm{rad}}$ and dispersions $\sigma$ of the different (outlier cleaned) GC/UCD populations (cf. Fig.~\ref{fig:velcolhist} and Fig.~\ref{fig:magdisp}). $\sigma_{\mathrm{PM}}$ is the dispersion from the \citet{1993ASPC...50..357P} estimator. The dividing magnitude for the faint and the bright sample is $M_V=-10.75$~mag.}
	\label{tab:velcolhist}
	\centering	
		\begin{tabular}{l c l l l}
		\hline\hline
		Population & $N$ & $\bar{v}_{\mathrm{rad}}$ & $\sigma$ & $\sigma_{\mathrm{PM}}$ \\
		~ &  & [\kms] & [\kms] & [\kms] \\
		\hline
		All                 & 118 & $3717 \pm 65$  & $710 \pm 46$  & $706 \pm 46$\\
		All$^*$             & 113 & $3715 \pm 56$  & $600 \pm 40$  & $596 \pm 40$\\
		All, $R<5'$         & 87  & $3715 \pm 77$  & $718 \pm 54$  & $714 \pm 55$\\
		All$^*$, $R<5'$     & 85  & $3770 \pm 69$  & $632 \pm 49$  & $629 \pm 49$\\
		All, $R\geq5'$      & 31  & $3720 \pm 123$ & $686 \pm 87$  & $682 \pm 88$\\
		All$^*$, $R\geq5'$  & 28  & $3548 \pm 85$  & $450 \pm 60$  & $444 \pm 61$\\
		Red                 & 104 & $3698 \pm 69$  & $705 \pm 49$  & $701 \pm 49$\\
		Red$^*$             & 100 & $3716 \pm 61$  & $611 \pm 43$  & $608 \pm 44$\\
		Red, $R<5'$         & 78  & $3702 \pm 83$  & $735 \pm 59$  & $731 \pm 59$\\
		Red$^*$, $R<5'$     & 76  & $3762 \pm 74$  & $642 \pm 52$  & $639 \pm 52$\\
	    Red, $R\geq5'$      & 26  & $3685 \pm 119$ & $607 \pm 84$  & $603 \pm 85$\\
		Red$^*$, $R\geq5'$  & 24  & $3569 \pm 96$  & $471 \pm 68$  & $465 \pm 69$\\
		Blue                & 14  & $3856 \pm 195$ & $729 \pm 138$ & $726 \pm 138$\\
		Blue$^*$            & 13  & $3706 \pm 141$ & $507 \pm 99$  & $502 \pm 100$\\
		Blue, $R<5'$        & 9   & $3830 \pm 180$ & $539 \pm 127$ & $535 \pm 128$\\
		Blue, $R\geq5'$     & 5   & $3902 \pm 439$ & $981 \pm 310$ & $920 \pm 268$\\
		Blue$^*$, $R\geq5'$ & 4   & $3425 \pm 129$ & $257 \pm 91$  & $247 \pm 94$\\
		\hline
		Faint               & 58  & $3697 \pm 107$ & $817 \pm 76$ & ... \\
		Bright              & 60  & $3735 \pm 76$  & $588 \pm 54$ & ... \\
		Faint$^*$           & 55  & $3755 \pm 93$  & $689 \pm 66$ & ... \\
		Bright$^*$          & 58  & $3676 \pm 65$  & $499 \pm 46$ & ... \\
		Faint, $R<5'$       & 49  & $3710 \pm 119$ & $833 \pm 84$ & ... \\
		Bright, $R<5'$      & 38  & $3722 \pm 87$  & $535 \pm 61$ & ... \\
		Faint$^*$, $R<5'$   & 47  & $3808 \pm 102$ & $699 \pm 72$ & ... \\
		Bright$^*$, $R<5'$  & 38  & $3722 \pm 87$  & $535 \pm 61$ & ... \\
		\hline
		\end{tabular}
		\tablefoot{\tablefoottext{*}{Sample after applying the outlier rejection method (Sect.~\ref{sec:kinematics}).}}
\end{table}

\subsubsection{Velocity dispersions}
\label{sec:dispersions}
In most of the cases, the two velocity dispersions ($\sigma$ and $\sigma_{\mathrm{PM}}$) given in Table~\ref{tab:velcolhist} do not differ from each other by more than 6~\kms. The velocity dispersions of both the inner and the outer populations (without outlier rejection) are with $\sim 700$~\kms~comparable to what \citet{2003ApJ...591..764C} found for the Hydra\,I cluster velocity dispersion, i.e. $\sigma=724\pm31$~\kms. Naturally, the velocity dispersions of the outlier-cleaned samples are always lower than those of the full samples. However, removing the three objects above $5000$~\kms~and with distances $80<R<150$~kpc (see Fig.~\ref{fig:radialdist}) would lead to a considerable drop in the velocity dispersion from $\sim 700$~\kms~to $\sim 450$~\kms~(cf. Table~\ref{tab:velcolhist}).

Assuming that most of the GCs/UCDs at distances larger than $70$~kpc dynamically belong to NGC~3311 (see Sect.~\ref{sec:meanvelocities}), a lower velocity dispersion compared to the overall cluster galaxy velocity dispersion might be expected. This has for example been observed for a sample of UCDs in the Coma cluster core \citep{2010arXiv1009.3950C}. However, if including the three questionable objects, the velocity dispersions of the outer population is comparable to the one of the inner population. Those objects might also belong to a different cluster galaxy with a high radial velocity ($\gtrsim 5000$~\kms), hence causing the high velocity dispersion in the outer sample, but none of the three questionable objects is located close to a major cluster galaxy with a similar high radial velocity. On the other hand, if the majority of the outer GCs/UCDs belongs to an intra-cluster population, one would expect the velocity dispersion to be of similar order to the velocity dispersion of the cluster galaxies, which is about $800$~\kms~at $100$~kpc \citep{2006MNRAS.367.1463L}. However, the GC/UCD velocity dispersion at distances larger than $70$~kpc is statistically less well defined than for smaller distances, thus we cannot favour one of these scenarios. This holds for the full sample as well as for the red sub-sample. Due to the low number of objects, the values for the blue sub-populations are highly uncertain and do not allow a deeper analysis. The radial velocity dispersion profile of the GC/UCD system and its significance for the dynamics of NGC~3311 is discussed in a parallel contribution \citep{2011arXiv1103.2053R}.

\begin{figure}
	\resizebox{\hsize}{!}{\includegraphics{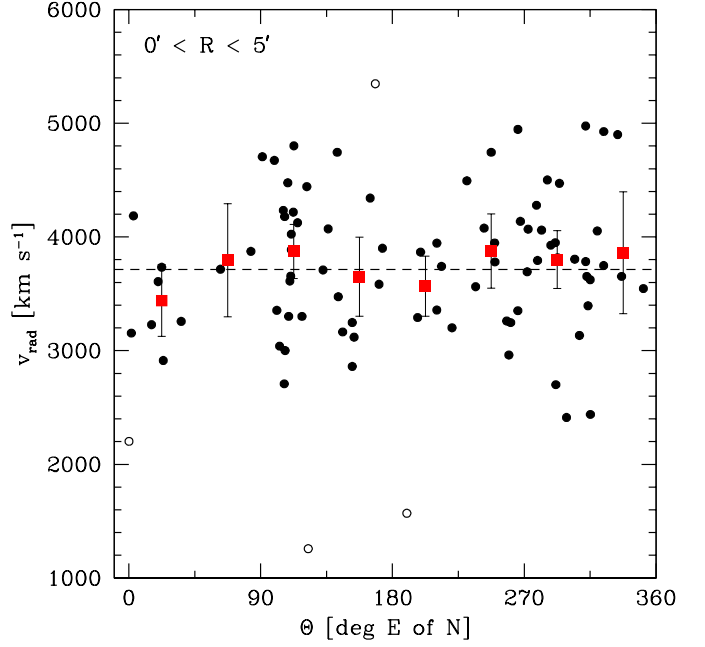}}
	\caption{Radial velocity $v_{\mathrm{rad}}$ as a function of azimuthal position $\Theta$ measured in degrees East of North on the projected sky. Filled circles are all objects within a projected radial distance of $R=5\arcmin$ to NGC~3311. Open circles are objects that deviate more than $2\sigma$ from the mean radial velocity of the sample (dashed line) The red filled squares give the mean radial velocity of the objects in bins of 45 degrees. The error bars represent the 90\% confidence interval of the mean.}
	\label{fig:rotation}
\end{figure}

In Fig.~\ref{fig:magdisp} we compare the velocity dispersion $\sigma$ of bright and faint objects by splitting the GC/UCD sample at six different magnitudes around $M_V=-11$~mag. This is done for both the full sample and the outlier-cleaned sample. Additionally, we do the same analysis only for objects within $5\arcmin$ around NGC~3311 (see the lower two panels of Fig.~\ref{fig:magdisp}).

In both the full and the outlier-cleaned sample, bright GCs/UCDs have a lower velocity dispersion than fainter objects. This is true for dividing magnitudes $M_V \lesssim -10.5$~mag, and it is most pronounced at $M_V = -10.75$~mag, at which also the sample sizes are almost identical. At this magnitude, the $\sigma$-values differ by more than 200~\kms~($\sim 3\sigma$ significance), and up to 300~\kms~($\sim 4\sigma$ significance) for objects with $R<5\arcmin$ (see also Table~\ref{tab:velcolhist}). The differences for the outlier-cleaned samples (full and inner sample) are smaller, but still more than $160$~\kms. At fainter dividing magnitudes the differences are not any longer significant. 

At first view, the lower velocity dispersion of the bright sample might be explained by the selection effects described above (Sect.~\ref{sec:kinematics}): due to the configuration of the VIMOS pointings (Fig.~\ref{fig:fields}), bright objects are preferentially probed at larger projected distances to NGC~3311. At the same time, these objects exhibit a lower velocity dispersion, as can be seen in Fig.~\ref{fig:radialdist} and Table~\ref{tab:velcolhist}. However, when only regarding objects within $5\arcmin$ around NGC~3311, the difference in $\sigma$ is even larger and cannot only be explained by selection effects. We discuss the implications of this finding in more detail in Sect.~\ref{sec:conclusions}.

\subsubsection{Rotation}
In order to further investigate the dynamical properties of the GC system around NGC~3311, we restrict our sample to objects within a projected distance of $5\arcmin$ and with radial velocities $\pm 2 \sigma$ around the mean of this sample. We then plot their radial velocity as a function of their azimuthal position in the projected sky (Fig.~\ref{fig:rotation}). If the GC system was rotating as a whole, we expect in such a diagram a sine pattern of the form 
\begin{equation}
v_{\mathrm{rad}} (\Theta) = v_{\mathrm{sys}} + A_{\mathrm{rot}} \cdot \sin(\Theta - \Theta_0),
\end{equation}
with $v_{\mathrm{sys}}$ being the mean radial velocity of the sample, $A_{\mathrm{rot}}$ the rotation amplitude and $\Theta_0$ the projected rotation axis. It turns out that a meaningful fit to the data is not possible, neither to the individual data points nor to the binned data. The small sample size (only 83 objects) and a significant incompleteness of the sample at $\Theta \sim 65\degr$ and $\Theta \sim 200 \degr$ prevents a deeper analysis.

\section{Discussion and conclusions}
\label{sec:conclusions}
With two extensive spectroscopic surveys we confirmed that the core of the Hydra\,I galaxy cluster contains a large population of at least 50 UCDs, as it has been presumed on the basis of earlier photometric studies \citep{2007ApJ...668L..35W, 2008ApJ...681.1233W}. The presence of the very pronounced diffuse light component of NGC~3311 implies that environmental effects have been very important in shaping the central region of the Hydra\,I cluster, probably more than in the case of the Fornax, Virgo and Centaurus clusters. The dust structure in the center of NGC~3311 is a clear evidence for a recent interaction with a gas-rich galaxy (see e.g. Fig.~\ref{fig:n3311core} in the appendix and \citealt{1994AJ....108..102G}). Together with the large number of UCDs, this strengthens the idea of an interaction driven UCD formation process in Hydra\,I, in which several UCDs are the remnant nuclei of dwarf galaxies whose stellar envelopes were stripped off during interaction with their host galaxy or galaxy cluster \citep{2003MNRAS.344..399B}. 

This hypothesis is supported by the similar luminosities, colours and sizes of UCDs and nuclei of dwarf galaxies \citep[e.g.][]{2006ApJS..165...57C}. Many of the confirmed Hydra\,I UCDs have much brighter magnitudes than UCD candidates previously identified in photometric studies (see Sect.~\ref{sec:photprops}). While the latter might rather represent the bright end of the globular cluster luminosity function (cf. Fig.~\ref{fig:wehnercmd}), some of the brighter objects might be nuclei of stripped dwarf galaxies. At magnitudes brighter than $M_V\sim -12$~mag (see Figs.~\ref{fig:cmd} and \ref{fig:hstcmd}), we find objects extending the red GC population towards higher luminosities, as well as objects with rather blue colours, which coincide with nuclei of dwarf galaxies in the colour-magnitude diagram. 

\citet{2011arXiv1102.0001N} suggested a general scheme of GC/UCD formation and the connection to galaxy nuclei \citep[see also][]{2011A&A...525A..86D, 2011MNRAS.tmp...66C}. In this picture, objects below $M_V \sim -10$~mag are "normal" GCs (red and blue) with a common mean size, plus a fraction of low-mass nuclei, indistinguishable from GCs. For magnitudes $-10 \gtrsim M_V \gtrsim -13$~mag , a mass-size and a mass-metallicity relation (the "blue tilt") is observed \citep[e.g.][]{2005ApJ...627..203H, 2010ApJ...710.1672M}. UCDs (or giant globular clusters), extending the globular cluster luminosity function, as well as giant and dwarf nuclei populate this luminosity regime, the latter having on average bluer colours. In Hydra\,I, HUCD9, HUCD18 and HUCD35, all having rather blue colours, might be examples of such stripped dwarf galaxy nuclei. Even brighter (more massive) objects are expected to be exclusively stripped nuclei. HUCD1 is the only object in our sample that meets this criterion. For the case of Hydra\,I, wide-field photometry, deeper and more accurate than the VIMOS photometry, would be required to perform a robust investigation of the practicability of this classification scheme.

The above assumption of the brightest UCDs as descendants of nucleated dwarf galaxies is supported by the finding that the velocity dispersion of bright objects is smaller than that of fainter objects (see Sect.~\ref{sec:dispersions}, Table~\ref{tab:velcolhist} and Fig.~\ref{fig:magdisp}). Assuming that several passages through the cluster center are necessary for nucleated dwarf galaxies to loose their stellar envelopes, they might end up on more circular orbits, leading to a lower velocity dispersion. This might explain the difference in velocity dispersion between bright UCDs, potentially being the remnant nuclei of dwarf galaxies, and fainter objects, belonging to the globular cluster system of the central galaxy.

An alternative explanation for the difference in velocity dispersion may be that the population of bright GCs/UCDs exhibits a steeper number density profile than the fainter population. This would be plausible in a scenario, where the brightest clusters formed in violent starbursts, which naturally occurred with higher probability near the center of the galaxy cluster than at large radii. The different velocity dispersions would then be a simple consequence of the Jeans-equation, like the different dispersions of blue and red GCs in NGC~1399 \citep{2004AJ....127.2094R, 2010A&A...513A..52S}. A change in the number density profile would not be visible in the distribution of our slits, which do not represent the true spatial distribution of the objects. A complete photometric sample over a larger radial range would be required for this kind of analysis. The study of \citet{2008ApJ...681.1233W} reaches out to only $3\arcmin$, and the authors do not comment on possible differences in the number density profiles of bright and faint objects.  

Another, related view is that the dispersion of the fainter population is boosted by objects on highly elongated orbits with large clustercentric distances. In the NGC~1399 system, one finds objects with very high individual velocities, suggesting apogalactic distances as high as 400~kpc, which may dynamically be assigned to the entire cluster rather than to the central galaxy \citep{2004AJ....127.2094R, 2010A&A...513A..52S}. Also the Hydra\,I data show that the objects with the highest individual velocities ($\vrad \gtrsim 4900$~\kms) tend to be faint. However, more radial velocities would be necessary for a more detailed interpretation.

\begin{acknowledgements}
The data published in this paper have been reduced using VIPGI, designed by the VIMOS Consortium and developed by INAF Milano. We thank Elizabeth Wehner for providing the list of photometrically identified candidate GCs/UCDs around NGC~3311. IM acknowledges support through DFG grant BE1091/13-1. TR acknowledges financial support from the Chilean Center for Astrophysics, FONDAP Nr.~15010003, from FONDECYT project Nr.~1100620, and from the BASAL Centro de Astrofisica y Tecnologias Afines (CATA) PFB-06/2007.
\end{acknowledgements}

\bibliographystyle{aa}
\bibliography{hydraucds}

\Online

\begin{appendix}
\section{~}

\begin{figure}[hb!]
	\resizebox{\hsize}{!}{\includegraphics{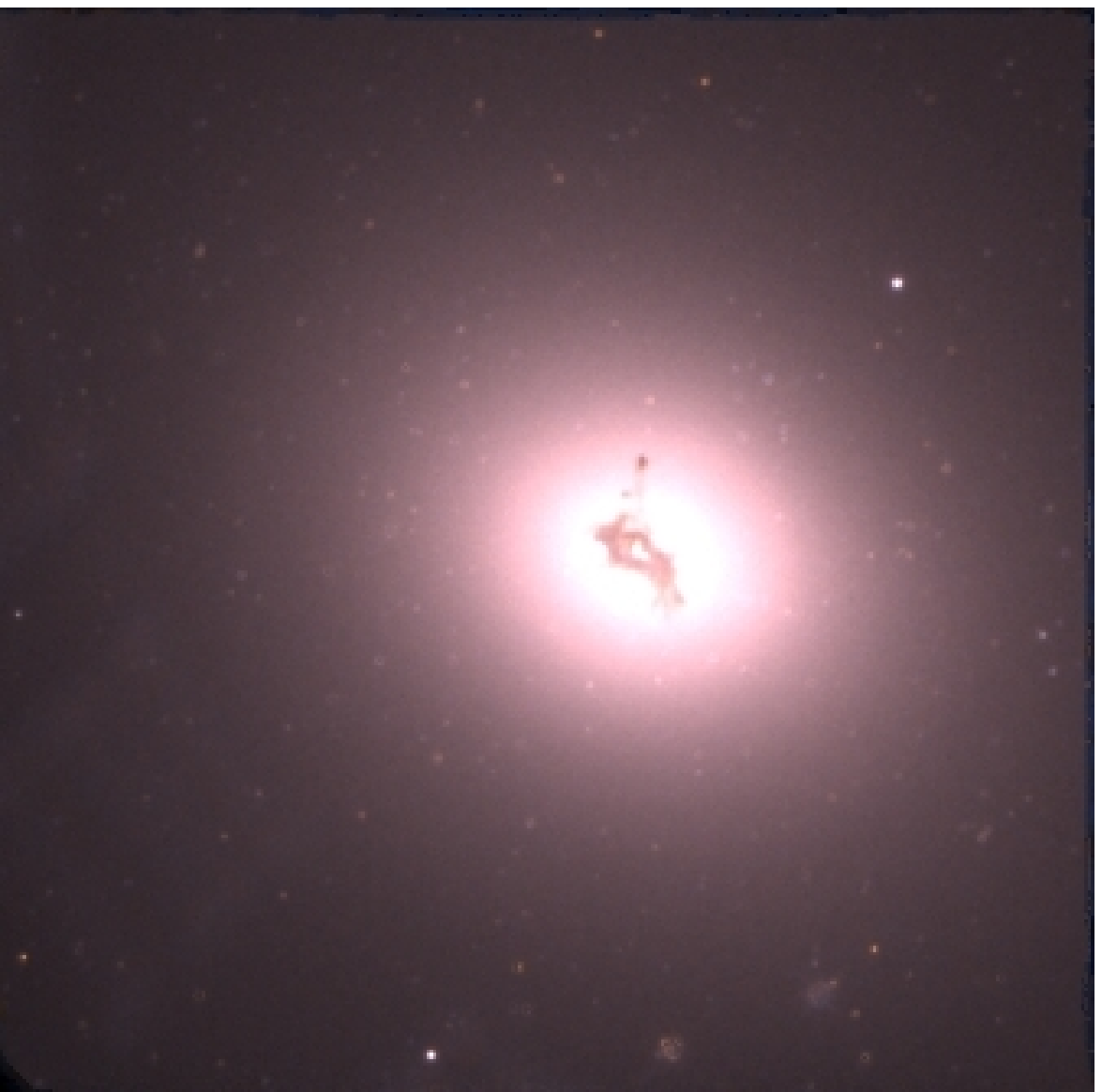}}
	\caption{Colour composite image of the dust lane in the core of NGC~3311 with HST/WFPC2 $F555W$ and $F814W$ filters. The field of view is $\sim 35 \times 35$~arcsec. }
	\label{fig:n3311core}
\end{figure}

\onecolumn

\longtab{1}{
\begin{longtable}{rrrrccc}
\caption{\label{tab:hydraucdsgcs} Catalogue of the 118 identified cluster GCs/UCDs. The first column gives the object ID, right ascension and declination (J2000.0) are given in columns two and three. The fourth column lists $R$, the projected distance in arcmin to NGC~3311. Columns 5 and 6 list the extinction corrected apparent magnitude $V_0$ and the colour $(V-R)_0$. The last column gives the radial velocity $\vrad$ with the according error. The table is ordered by increasing apparent magnitude.}\\
\hline\hline
ID & $\alpha$(2000.0) & $\delta$(2000.0) & $R$ & $V_0$ & $(V-R)_0$ & $\vrad$ \\
~ & [h:m:s] & [$^\circ$:$\arcmin$:$\arcsec$] & [arcmin] & [mag] & [mag] & [\kms] \\
\hline
\endfirsthead
\caption{continued.}\\
\hline\hline
ID & $\alpha$(2000.0) & $\delta$(2000.0) & $R$ & $V_0$ & $(V-R)_0$ & $\vrad$ \\
~ & [h:m:s] & [$^\circ$:$\arcmin$:$\arcsec$] & [arcmin] & [mag] & [mag] & [\kms] \\
\hline
\endhead
\hline
\endfoot

HUCD1 & 10:36:42.5 & -27:32:52.9 & $1.19$ & $20.03$ & $0.60$ & $ 3352 \pm  42$ \\ 
HUCD2 & 10:36:19.4 & -27:26:05.9 & $8.09$ & $20.65$ & $0.61$ & $ 3547 \pm  52$ \\ 
HUCD3 & 10:36:41.9 & -27:19:44.8 & $11.96$ & $20.85$ & $0.52$ & $ 3354 \pm  56$ \\ 
HUCD4 & 10:36:34.8 & -27:34:45.8 & $3.66$ & $20.86$ & $0.64$ & $ 3563 \pm  43$ \\ 
HUCD5 & 10:36:34.4 & -27:30:15.3 & $2.55$ & $21.06$ & $0.61$ & $ 3804 \pm  61$ \\ 
HUCD6 & 10:36:43.9 & -27:37:57.3 & $6.26$ & $21.07$ & $0.53$ & $ 3980 \pm  50$ \\ 
HUCD7 & 10:36:42.1 & -27:24:03.5 & $7.64$ & $21.15$ & $0.53$ & $ 3986 \pm  43$ \\ 
HUCD8 & 10:36:43.8 & -27:32:22.2 & $0.71$ & $21.22$ & $0.54$ & $ 3301 \pm  53$ \\ 
HUCD9 & 10:35:36.7 & -27:17:40.2 & $21.67$ & $21.24$ & $0.47$ & $ 3137 \pm  72$ \\ 
HUCD10 & 10:36:31.8 & -27:29:13.8 & $3.70$ & $21.30$ & $0.55$ & $ 4976 \pm  46$ \\ 
HUCD11 & 10:36:32.4 & -27:29:22.1 & $3.49$ & $21.30$ & $0.60$ & $ 3784 \pm  49$ \\ 
HUCD12 & 10:36:45.5 & -27:29:52.2 & $1.95$ & $21.33$ & $0.58$ & $ 3608 \pm  99$ \\ 
HUCD13 & 10:36:01.9 & -27:23:01.3 & $13.42$ & $21.42$ & $0.52$ & $ 4254 \pm  86$ \\ 
HUCD14 & 10:36:31.4 & -27:30:25.7 & $3.12$ & $21.44$ & $0.60$ & $ 4472 \pm  57$ \\ 
HUCD15 & 10:36:49.3 & -27:36:01.9 & $4.63$ & $21.44$ & $0.59$ & $ 3655 \pm  60$ \\ 
HUCD16 & 10:36:41.1 & -27:31:18.0 & $0.59$ & $21.61$ & $0.55$ & $ 3653 \pm  48$ \\ 
HUCD17 & 10:36:06.0 & -27:36:33.3 & $10.40$ & $21.66$ & $0.62$ & $ 5082 \pm  43$ \\ 
HUCD18 & 10:36:43.5 & -27:31:49.4 & $0.22$ & $21.67$ & $0.66$ & $ 3165 \pm  63$ \\ 
HUCD19 & 10:36:32.0 & -27:30:12.7 & $3.08$ & $21.67$ & $0.62$ & $ 2412 \pm  71$ \\ 
HUCD20 & 10:36:14.0 & -27:27:56.4 & $8.13$ & $21.71$ & $0.64$ & $ 3124 \pm  74$ \\ 
HUCD21 & 10:36:47.4 & -27:31:06.6 & $1.28$ & $21.71$ & $0.60$ & $ 3716 \pm  43$ \\ 
HUCD22 & 10:36:58.0 & -27:35:58.9 & $5.72$ & $21.73$ & $0.60$ & $ 3127 \pm  77$ \\ 
HUCD23 & 10:37:43.1 & -27:36:00.9 & $15.67$ & $21.74$ & $0.58$ & $ 3760 \pm  40$ \\ 
HUCD24 & 10:36:56.8 & -27:33:30.7 & $3.93$ & $21.75$ & $0.55$ & $ 2861 \pm  72$ \\ 
HUCD25 & 10:36:44.8 & -27:34:15.6 & $2.61$ & $21.76$ & $0.66$ & $ 3354 \pm  37$ \\ 
HUCD26  & 10:36:35.0 & -27:32:59.9 & $2.35$ & $21.82$ & $0.56$ & $ 3742 \pm  96$ \\ 
HUCD27  & 10:36:35.0 & -27:29:44.4 & $2.77$ & $21.88$ & $0.58$ & $ 3624 \pm  61$ \\ 
HUCD28 & 10:36:52.9 & -27:24:33.1 & $7.58$ & $21.88$ & $0.53$ & $ 2664 \pm  91$ \\ 
HUCD29 & 10:36:48.2 & -27:22:18.5 & $9.49$ & $21.89$ & $0.60$ & $ 3301 \pm  39$ \\ 
HUCD30 & 10:36:50.0 & -27:31:54.9 & $1.81$ & $21.91$ & $0.61$ & $ 3901 \pm  48$ \\ 
HUCD31 & 10:36:05.6 & -27:27:04.0 & $10.39$ & $21.92$ & $0.52$ & $ 3725 \pm  62$ \\ 
HUCD32 & 10:36:57.3 & -27:26:01.1 & $6.74$ & $21.99$ & $0.48$ & $ 3294 \pm  60$ \\ 
HUCD33 & 10:36:43.4 & -27:25:20.9 & $6.35$ & $22.01$ & $0.56$ & $ 4217 \pm  61$ \\ 
HUCD34 & 10:36:48.5 & -27:34:43.4 & $3.34$ & $22.04$ & $0.49$ & $ 4125 \pm  52$ \\ 
HUCD35 & 10:36:31.7 & -27:30:32.0 & $3.01$ & $22.04$ & $0.56$ & $ 3818 \pm  59$ \\ 
HUCD36 & 10:37:01.2 & -27:27:19.5 & $6.35$ & $22.06$ & $0.45$ & $ 5810 \pm  70$ \\ 
HUCD37 & 10:36:35.6 & -27:35:11.0 & $3.92$ & $22.08$ & $0.63$ & $ 4078 \pm  35$ \\ 
HUCD38 & 10:36:07.2 & -27:32:51.4 & $8.97$ & $22.08$ & $0.57$ & $ 3821 \pm  89$ \\ 
HUCD39 & 10:36:45.1 & -27:31:51.1 & $0.58$ & $22.12$ & $0.59$ & $ 4342 \pm  60$ \\ 
HUCD40 & 10:36:39.4 & -27:30:53.6 & $1.17$ & $22.14$ & $0.63$ & $ 3395 \pm  70$ \\ 
HUCD41 & 10:36:37.8 & -27:34:44.2 & $3.29$ & $22.15$ & $0.54$ & $ 4745 \pm  65$ \\ 
HUCD42 & 10:36:36.3 & -27:32:16.3 & $1.74$ & $22.16$ & $0.55$ & $ 3867 \pm  53$ \\ 
HUCD43 & 10:36:48.4 & -27:35:03.1 & $3.63$ & $22.18$ & $0.60$ & $ 4802 \pm  46$ \\ 
HUCD44 & 10:36:48.1 & -27:32:23.3 & $1.50$ & $22.23$ & $0.58$ & $ 3248 \pm  51$ \\ 
HUCD45 & 10:36:50.2 & -27:36:10.3 & $4.83$ & $22.24$ & $0.56$ & $ 4218 \pm  87$ \\ 
HUCD46 & 10:36:33.2 & -27:30:54.8 & $2.53$ & $22.26$ & $0.55$ & $ 3927 \pm  61$ \\ 
HUCD47 & 10:36:50.7 & -27:32:01.0 & $1.99$ & $22.28$ & $0.48$ & $ 3584 \pm  68$ \\ 
HUCD48 & 10:36:40.6 & -27:26:27.3 & $5.27$ & $22.29$ & $0.59$ & $ 3850 \pm  56$ \\ 
HUCD49 & 10:36:46.3 & -27:34:18.6 & $2.75$ & $22.34$ & $0.54$ & $ 4477 \pm  68$ \\ 
HUCD50 & 10:36:28.1 & -27:31:06.7 & $3.74$ & $22.34$ & $0.54$ & $ 3794 \pm  78$ \\ 
HUCD51 & 10:35:45.7 & -27:33:38.4 & $14.41$ & $22.35$ & $0.64$ & $ 4079 \pm  66$ \\ 
HUCD52 & 10:35:52.9 & -27:35:32.3 & $13.07$ & $22.37$ & $0.63$ & $ 3596 \pm  69$ \\ 
53 & 10:36:39.5 & -27:29:48.3 & $2.07$ & $22.38$ & $0.46$ & $ 3653 \pm  59$ \\ 
54 & 10:36:44.5 & -27:30:44.7 & $1.04$ & $22.43$ & $0.55$ & $ 2914 \pm  53$ \\ 
55 & 10:36:45.3 & -27:29:26.5 & $2.34$ & $22.47$ & $0.57$ & $ 3229 \pm  80$ \\ 
56 & 10:36:40.9 & -27:25:09.0 & $6.57$ & $22.49$ & $0.44$ & $ 3439 \pm  61$ \\ 
57 & 10:36:30.1 & -27:34:26.3 & $4.20$ & $22.56$ & $0.62$ & $ 3201 \pm  81$ \\ 
58 & 10:36:46.4 & -27:29:32.1 & $2.34$ & $22.57$ & $0.40$ & $ 3734 \pm  71$ \\ 
59 & 10:36:38.3 & -27:32:21.4 & $1.30$ & $22.58$ & $0.61$ & $ 3358 \pm  69$ \\ 
60 & 10:36:50.0 & -27:38:27.5 & $6.99$ & $22.61$ & $0.62$ & $ 3532 \pm  57$ \\ 
61 & 10:36:50.0 & -27:32:35.0 & $2.00$ & $22.63$ & $0.57$ & $ 3119 \pm  78$ \\ 
62 & 10:36:43.8 & -27:33:13.0 & $1.54$ & $22.64$ & $0.62$ & $ 4674 \pm  64$ \\ 
63 & 10:36:43.8 & -27:27:12.3 & $4.50$ & $22.65$ & $0.59$ & $ 4185 \pm  51$ \\ 
64 & 10:36:38.0 & -27:30:16.6 & $1.86$ & $22.66$ & $0.55$ & $ 4053 \pm  71$ \\ 
65 & 10:36:53.1 & -27:35:43.6 & $4.77$ & $22.67$ & $0.55$ & $ 1259 \pm  91$ \\ 
66 & 10:36:39.1 & -27:31:20.1 & $1.01$ & $22.71$ & $0.70$ & $ 3950 \pm  55$ \\ 
67 & 10:36:45.3 & -27:33:43.9 & $2.12$ & $22.80$ & $0.62$ & $ 3001 \pm  83$ \\ 
68 & 10:36:39.5 & -27:31:28.1 & $0.85$ & $22.81$ & $0.63$ & $ 4502 \pm  54$ \\ 
69 & 10:36:41.2 & -27:32:46.7 & $1.15$ & $22.82$ & $0.58$ & $ 3780 \pm  47$ \\ 
70 & 10:36:44.9 & -27:33:36.4 & $1.98$ & $22.84$ & $0.49$ & $ 4234 \pm  54$ \\ 
71 & 10:37:01.5 & -27:24:07.4 & $8.91$ & $22.86$ & $0.50$ & $ 2874 \pm  51$ \\ 
72 & 10:36:47.0 & -27:35:18.1 & $3.75$ & $22.86$ & $0.51$ & $ 2708 \pm  79$ \\ 
73 & 10:36:38.8 & -27:30:55.0 & $1.28$ & $22.86$ & $0.58$ & $ 3134 \pm  76$ \\ 
74 & 10:36:26.8 & -27:32:23.5 & $4.06$ & $22.90$ & $0.55$ & $ 1570 \pm  86$ \\ 
75 & 10:36:34.2 & -27:21:32.3 & $10.39$ & $22.91$ & $0.65$ & $ 2755 \pm  61$ \\ 
76 & 10:36:45.8 & -27:33:43.9 & $2.16$ & $22.92$ & $0.59$ & $ 3612 \pm  47$ \\ 
77 & 10:36:43.2 & -27:28:27.6 & $3.24$ & $22.92$ & $0.57$ & $ 3154 \pm  69$ \\ 
78 & 10:35:53.0 & -27:23:06.7 & $15.14$ & $22.96$ & $0.43$ & $ 3830 \pm  68$ \\ 
79 & 10:36:30.6 & -27:21:37.1 & $10.54$ & $22.96$ & $0.61$ & $ 3912 \pm  52$ \\ 
80 & 10:36:43.0 & -27:34:09.3 & $2.46$ & $23.00$ & $0.60$ & $ 4706 \pm  58$ \\ 
81 & 10:37:03.7 & -27:26:42.0 & $7.23$ & $23.00$ & $0.54$ & $ 3659 \pm  59$ \\ 
82 & 10:36:42.9 & -27:27:59.1 & $3.72$ & $23.01$ & $0.55$ & $ 2202 \pm  51$ \\ 
83 & 10:36:49.9 & -27:33:02.0 & $2.22$ & $23.04$ & $0.58$ & $ 3474 \pm  56$ \\ 
84 & 10:36:40.5 & -27:35:12.2 & $3.55$ & $23.06$ & $0.58$ & $ 3247 \pm  48$ \\ 
85 & 10:36:50.7 & -27:33:13.2 & $2.48$ & $23.08$ & $0.56$ & $ 4745 \pm  66$ \\ 
86 & 10:36:31.5 & -27:33:21.0 & $3.27$ & $23.09$ & $0.57$ & $ 3946 \pm  56$ \\ 
87 & 10:36:50.9 & -27:35:28.4 & $4.28$ & $23.10$ & $0.65$ & $ 3301 \pm  55$ \\ 
88 & 10:36:46.0 & -27:35:10.0 & $3.56$ & $23.10$ & $0.64$ & $ 3040 \pm  93$ \\ 
89 & 10:36:37.7 & -27:29:55.2 & $2.19$ & $23.10$ & $0.56$ & $ 4927 \pm  72$ \\ 
90 & 10:36:32.3 & -27:31:34.7 & $2.63$ & $23.11$ & $0.64$ & $ 4068 \pm  59$ \\ 
91 & 10:36:42.1 & -27:30:28.1 & $1.25$ & $23.14$ & $0.64$ & $ 3546 \pm  96$ \\ 
92 & 10:36:40.9 & -27:30:42.7 & $1.10$ & $23.16$ & $0.46$ & $ 4901 \pm  70$ \\ 
93 & 10:36:41.9 & -27:36:40.4 & $4.98$ & $23.17$ & $0.53$ & $ 4137 \pm  59$ \\ 
94 & 10:36:30.0 & -27:31:01.9 & $3.26$ & $23.19$ & $0.56$ & $ 4060 \pm  56$ \\ 
95 & 10:36:39.0 & -27:30:22.1 & $1.64$ & $23.21$ & $0.63$ & $ 3748 \pm  99$ \\ 
96 & 10:36:45.5 & -27:23:49.7 & $7.90$ & $23.22$ & $0.63$ & $ 3571 \pm  46$ \\ 
97 & 10:36:46.4 & -27:31:35.8 & $0.89$ & $23.28$ & $0.52$ & $ 3874 \pm  92$ \\ 
98 & 10:36:58.2 & -27:25:30.2 & $7.29$ & $23.28$ & $0.51$ & $ 5091 \pm  85$ \\ 
99 & 10:36:48.3 & -27:31:59.1 & $1.41$ & $23.28$ & $0.63$ & $ 5347 \pm  54$ \\ 
100 & 10:36:50.9 & -27:33:38.4 & $2.80$ & $23.29$ & $0.58$ & $ 4071 \pm  68$ \\ 
101 & 10:36:37.7 & -27:32:05.8 & $1.34$ & $23.29$ & $0.51$ & $ 3291 \pm  48$ \\ 
102 & 10:36:50.3 & -27:29:06.7 & $3.19$ & $23.30$ & $0.49$ & $ 3257 \pm  62$ \\ 
103 & 10:36:32.4 & -27:34:53.6 & $4.12$ & $23.31$ & $0.64$ & $ 4494 \pm  67$ \\ 
104 & 10:36:32.6 & -27:26:30.0 & $5.79$ & $23.33$ & $0.68$ & $ 4194 \pm  54$ \\ 
105 & 10:36:57.1 & -27:36:38.4 & $6.09$ & $23.34$ & $0.58$ & $ 2767 \pm  77$ \\ 
106 & 10:36:41.9 & -27:34:52.6 & $3.19$ & $23.34$ & $0.60$ & $ 4946 \pm  73$ \\ 
107 & 10:36:40.3 & -27:34:43.3 & $3.09$ & $23.41$ & $0.58$ & $ 3261 \pm  56$ \\ 
108 & 10:36:51.5 & -27:35:14.6 & $4.16$ & $23.42$ & $0.60$ & $ 4443 \pm  60$ \\ 
109 & 10:36:29.1 & -27:31:34.8 & $3.43$ & $23.44$ & $0.55$ & $ 3693 \pm  60$ \\ 
110 & 10:36:44.7 & -27:32:55.5 & $1.31$ & $23.47$ & $0.46$ & $ 4024 \pm  53$ \\ 
111 & 10:36:40.9 & -27:34:15.9 & $2.61$ & $23.52$ & $0.45$ & $ 2962 \pm  64$ \\ 
112 & 10:36:41.2 & -27:32:49.4 & $1.20$ & $23.53$ & $0.53$ & $ 3948 \pm  77$ \\ 
113 & 10:36:46.0 & -27:34:22.2 & $2.78$ & $23.55$ & $0.57$ & $ 4179 \pm  46$ \\ 
114 & 10:36:31.9 & -27:30:37.1 & $2.95$ & $23.57$ & $0.60$ & $ 2701 \pm  65$ \\ 
115 & 10:36:29.6 & -27:31:12.8 & $3.34$ & $23.59$ & $0.56$ & $ 4279 \pm  63$ \\ 
116 & 10:36:47.5 & -27:32:58.9 & $1.74$ & $23.62$ & $0.60$ & $ 3710 \pm  86$ \\ 
117 & 10:36:35.8 & -27:29:56.6 & $2.48$ & $23.63$ & $0.74$ & $ 2439 \pm  77$ \\ 
118 & 10:36:45.3 & -27:33:19.8 & $1.75$ & $23.70$ & $0.66$ & $ 3887 \pm  53$ \\

\end{longtable}
}

\end{appendix}

\end{document}